\documentclass[prf,reprint,superscriptaddress,amsmath,amssymb,aps,longbibliography]{revtex4-1}

\usepackage{graphicx}
\usepackage{dcolumn}
\usepackage{bm}

\usepackage{siunitx}
\usepackage{makecell}
\usepackage{mathrsfs}
\usepackage{booktabs}
\newcommand{\ve}[1]{\boldsymbol{#1}}

\newcommand{\Rep}{\ensuremath{\mathrm{Re}_p}}
\newcommand{\Ga}{\ensuremath{\mathrm{Ga}}}

\begin{document}

\preprint{APS/123-QED}

\title{Smart navigation of a gravity-driven glider with adjustable centre-of-mass} 

\author{X. Jiang}
\thanks{These authors contributed equally to this work}
\affiliation{
  AML, Department of Engineering Mechanics, Tsinghua University, 100084 Beijing, China
}

\author{J. Qiu}
\thanks{These authors contributed equally to this work}
\affiliation{
  Department of Physics, Gothenburg University, 41296 Gothenburg, Sweden
}

\author{K. Gustavsson}
\affiliation{
  Department of Physics, Gothenburg University, 41296 Gothenburg, Sweden
}

\author{B. Mehlig}
\affiliation{
  Department of Physics, Gothenburg University, 41296 Gothenburg, Sweden
}

\author{L. Zhao}
\email{Contact author: zhaolihao@tsinghua.edu.cn}
\affiliation{
  AML, Department of Engineering Mechanics, Tsinghua University, 100084 Beijing, China
}

\date{\today}

\begin{abstract}
Artificial gliders are designed to disperse as they settle through a fluid, requiring precise navigation to reach target locations. We show that a compact glider settling in a viscous fluid can navigate by dynamically adjusting its centre-of-mass. Using fully resolved direct numerical simulations (DNS) and reinforcement learning, we find two optimal navigation strategies that allow the glider to reach its target location accurately. These strategies depend sensitively on how the glider interacts with the surrounding fluid. The nature of this interaction changes as the particle Reynolds number $\Rep$ changes.
Our results explain how the optimal strategy depends on $\Rep$. At large $\Rep$, the glider learns to tumble rapidly by moving its centre-of-mass as its orientation changes. This generates a large horizontal inertial lift force, which allows the glider to travel far. At  small $\Rep$, by contrast, high viscosity hinders tumbling. In this case, the glider learns to adjust its centre-of-mass so that it settles with a steady, inclined orientation that results in a horizontal viscous force. The horizontal range is much smaller than for large Re$_p$, because this viscous force is much smaller than the inertial lift force at large Re$_p$. 
\end{abstract}

\maketitle

\onecolumngrid

\section{Introduction} \label{sec:level1}
Seeds and small organisms evolved the ability to glide through air without propulsion, allowing navigation and dispersion with high efficiency~\cite{yanoviak2005Directed,cummins2018Separated,augspurger1986Morphology}.
These examples  have inspired the concepts of small artificial gliders, which could form intelligent networks for environmental measurements in air or water~\cite{Kim2021,kahn1999Next,Lermusiaux2017,leonard2007collective}.
These tiny gliders, unlike conventional ones~\cite{Mitchell2013,leonard2007collective}, are designed with a focus on miniaturisation, low cost, and low energy consumption.

One important question is how such gliders can manoeuvre to reach a prescribed target location. This requires deep understanding of glider-fluid interactions, which vary significantly with the particle settling Reynolds number, $\Rep=v_{\rm s} a/\nu$.
Here, $v_{\rm s}$ is the settling speed of the glider, $a$ its size, and $\nu$ is the kinematic viscosity of  the fluid.
As $\Rep$ increases from zero, fluid inertia amplifies hydrodynamic forces and torques~\cite{happel1983Low,pierson2021Hydrodynamic}.
Non-spherical particles experience an additional fluid-inertia torque compared to spherical ones.
At small $\Rep$, this torque can be computed perturbatively and is proportional to $\Rep$~\cite{khayat1989Inertia,lopez2017Inertial,gustavsson2021Effect}.
In addition, a particle rotating relative to the fluid experiences a lift force perpendicular to its slip velocity~\cite{candelier2023second}.
At large $\Rep$, calculations of forces and torques rely on empirical models that assume a quasi-steady disturbance flow~\cite{ern2012WakeInduceda}. Using such models, Refs.~\cite{pesavento2004Falling,Andersen2005} analysed gliders settling at $\Rep\sim 10^3$. When the wake of the settling particle becomes unstable, complex falling patterns emerge, including oscillations, tumbling, and even chaotic dynamics~\cite{field1997Chaotic}.

A pioneering study~\cite{Paoletti2011} investigated the navigation of a two-dimensional elliptical glider using an empirical model at $\Rep\sim 10^3$, representing a centimeter-scale glider in air. The authors showed that external control torque enables navigation through two strategies: tumbling and inclined settling, both generating a horizontal force via fluid-solid interaction.

Recent advances using reinforcement learning~\cite{Mnih2015,mehlig2021machine} have significantly improved the finding and understanding of optimal navigation strategies for smart particles that can sense their environment and adapt their behaviour in viscous flow~\cite{Colabrese2017,Novati2019,Gunnarson2021,Alageshan2020}.
Ref~\cite{Novati2019} applied reinforcement learning to the model in Ref.~\cite{Paoletti2011}, showing that a smart glider, capable of dynamically adjusting its control torque, learns to navigate by either tumbling or inclined settling. The optimal strategy depends on the mass density and upon the shape of the glider. Tumbling outperforms inclined settling for heavier or less elongated gliders~\cite{Novati2019}.

These studies highlight the potential of designing smart artificial gliders, but two key challenges remain. First, how can the control torque be applied? Gliders used in typical applications~\cite{Kim2021,kahn1999Next,Lermusiaux2017} are too small for propellers, so controlling such gliders like a drone is not an option. While magnetic or electric fields can control smart particles~\cite{jiang2022Controla}, they require external fields along the glider trajectory, limiting autonomy.
Second, how do the navigation strategies change with $\Rep$? $\Rep$ varies greatly in applications, and laboratory experiments are usually easier to perform when the glider settles slowly in a highly viscous fluid~\cite{lopez2017Inertial,Roy2019}. This yields a much smaller $\Rep$ than the values $\Rep\sim 10^3$ studied previously~\cite{Paoletti2011,Novati2019}.

\begin{figure}
  \includegraphics[width=8cm]{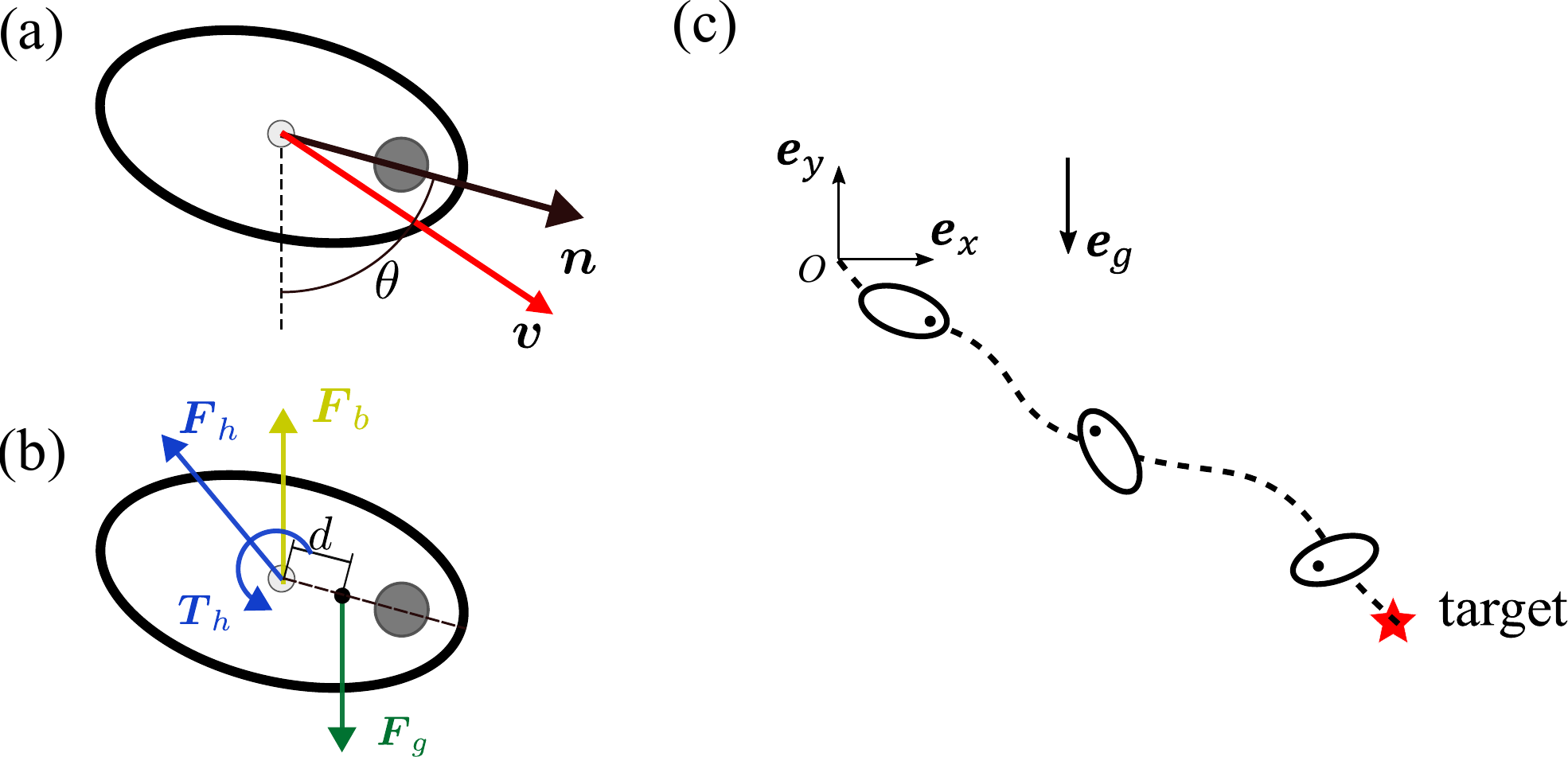}
  \caption{\label{fig1} (a) Glider with a movable mass (\raisebox{-0.5mm}{\Large$\bullet$}). Here $\boldsymbol{n}$ is  the unit vector along the major axis of the glider, $\theta$ is the tilt angle, and $\ve v$ is the velocity of the geometric centre. (b) Illustration of the forces and torques on a glider. The torques are evaluated with respect to the geometric centre. Buoyancy $\ve F_{\rm b}$ ($\circ$) induces no torque, while gravity $\ve F_{\rm g}$ (\raisebox{0.2mm}{\scriptsize$\bullet$}), causes the torque $T_g=-d\sin\theta$ because the centre-of-mass is located at distance $d$ along $\ve n$ from the geometric centre. Hydrodynamic forces on the glider surface result in a net force $\ve F_{\rm h}$ and torque $T_{\rm h}$ with respect to the geometric centre. (c) Schematic of active control of a settling glider in a quiescent fluid by adjusting its centre-of-mass. The red star represents the target point.}
\end{figure}

Here we address these two questions. First, we explore the possibility to control the glider using an adjustable 
centre-of-mass (Fig.~\ref{fig1}), inspired by  particles with asymmetric mass distributions that exhibit complex settling dynamics in a viscous fluid~\cite{Roy2019,jiangSettling2024}.
Second, we employ direct numerical simulations (DNS) to fully resolve particle-fluid interactions, in order to determine how the optimal strategy depends on Re$_p$. As an exemplary task, we consider a gravity-driven glider released in a quiescent fluid at different distances from the target.
Assuming the smart glider can measure its current phase-space configuration, we use reinforcement learning to determine the optimal strategy for adjusting its centre-of-mass to reach the target.

We find that the glider can exploit particle-fluid interactions to successfully navigate by changing its centre-of-mass. The optimal navigation strategy depends on $\Rep$: at small $\Rep$, the glider settles with steady  inclination, while it learns to tumble at larger values of $\Rep$.
Previous studies using empirical models for gliders at large particle Reynolds numbers ($\Rep\sim10^3$) identified the same two settling strategies~\cite{Paoletti2011,Novati2019}. Our work differs in three respects: we use DNS to investigate how the optimal strategy varies with $\Rep$; we implement an explicit control mechanism (moving centre-of-mass) rather than a prescribed torque; and we focus on gliders in viscous fluids, which are more practical for laboratory experiments.

\section{Methodology}
\subsection{Model}
Since the optimisation problem using DNS of particle-fluid interactions is computationally demanding, we follow~\cite{Paoletti2011, Novati2019} and consider the two-dimensional problem. 
We model the glider as a two-dimensional elliptical shell with aspect ratio $\lambda$ and a mass that can move along the direction $\ve n$ of its major axis [Fig. \ref{fig1} (a)]. 
Shifting this mass changes the centre-of-mass of the glider, enabling control.
Both the shell and the movable mass have equal linear mass density $\frac{1}{2}m_p$, where $m_p$ is the total mass per unit length.
We non-dimensionalise the problem using the semi-major axis length of the glider, $a$, the velocity scale $\sqrt{ag}$, and mass per unit length $m_{\rm p}$.
Here $g$ is the gravitational acceleration. The non-dimensional form of the governing equations of the glider reads

\begin{subequations}
  \label{eq:eom}
  \begin{align}
    \tfrac{{\rm d}}{{\rm d}t}\ve r & = \ve{v}\,\\
    \tfrac{{\rm d}^2}{{\rm d}t^2}(\ve r+\ve n d) & = \ve{F}_{\rm h}+\ve{F}_{\rm g}+\ve{F}_{\rm b}\,, 
    \label{eq::dynt}
    \\
    \big[\tfrac{1}{2}\tfrac{{\rm d}^2}{{\rm d}t^2}(\ve r+2\ve{n}d)\big] \cdot \ve{n} & = \tfrac{1}{2}\ve{F}_g \cdot \ve{n} + \ve{F}_{\rm s} \cdot \ve{n}\,.
    \label{eq::dynd} \\
    \tfrac{{\rm d}}{{\rm d}t}\theta = \omega\,, \quad\!\!\!
    \tfrac{{\rm d}}{{\rm d}t}(\omega J)  & = {T}_{\rm h} + T_g  - d(\ve n \wedge \tfrac{{\rm d}}{{\rm d}t}{\ve v})\cdot \ve e_z\,, 
    \label{eq::dynr}
  \end{align}
\end{subequations}
Here $\ve{r}$ and $\ve v$ are the position and velocity of the geometric centre of the glider, and dots denote time derivatives. In order to account for the moving mass, it is most convenient to write Newton's second law for the centre-of-mass, Eq.~(\ref{eq::dynt}). The left-hand side of Eq.~(\ref{eq::dynt}) is the acceleration of the centre-of-mass in terms of the position of the geometric centre $\ve r$, and the displacement $\ve nd$ between the geometric and mass centres.
The external forces on the glider are the hydrodynamic force $\ve F_{\rm h}$ exerted by the fluid upon the glider, gravity $\ve F_g=\ve{e}_g$, and buoyancy, $\ve F_b=-\beta^{-1}\ve{e}_g$, where $\ve e_g$ is the direction of gravity and $\beta = \rho_{\rm p}/\rho_{\rm f}$ is the mass-density ratio between the glider and fluid [Fig.~\ref{fig1} (b)].

When the glider shifts its centre-of-mass, momentum is conserved but redistributed between the movable mass, shell, and fluid via hydrodynamic interactions, generating transient forces and torques. 
To avoid infinite forces from instantaneous displacement, we model the connection between the movable mass and the shell as a damped spring, enabling continuous momentum transfer. 

The moving centre-of-mass adds a new dynamical degree of freedom: the centre-of-mass displacement $d$. Its dynamics is given by Eq.~(\ref{eq::dynd}).
The left-hand side is dimensionless mass times acceleration of the moving mass, located at $2\ve n d$ from the glider geometric centre, in the direction of $\ve n$. The right-hand side are the forces on the mass: gravity, $\tfrac{1}{2}\ve{F}_g$, and the damped spring force $\ve{F}_{\rm s} = [2k(d_e-d) - 2c\dot{d}]\ve{n}$. The spring stiffness $k=63.7$ and damping $c=9.0$ are chosen large to rapidly relax the centre-of-mass to its equilibrium distance $d_{\rm e}$.
The dynamics of the glider is controlled by adjusting $d_{\rm e}$.

Equation~(\ref{eq::dynr}) describes the angular dynamics of the glider around its geometric centre, with tilt angle $\theta$, and angular velocity $\omega$ around $\ve e_z$, the unit vector normal to the $x$-$y$ plane.
Equation~(\ref{eq::dynr}) is written in a co-moving frame that follows the translational motion of the geometric centre. The left-hand side of Eq.~(\ref{eq::dynr}) is the time derivative of angular momentum around the geometric centre, with moment of inertia $J=J_{\rm{shell}}+J_{\rm{mass}}+2d^2$. Here $J_{\rm{shell}}$ and $J_{\rm{mass}}$ are the moments of inertia of the elliptical shell and the circular movable mass around their respective centres, and the $2d^2$ contribution arises from the parallel axis theorem.
The hydrodynamic torque around the geometric centre is denoted by $T_{\rm h}$. The torque due to gravity, $T_g = -d \sin\theta$, arises due to the mass-centre displacement, see Fig.~\ref{fig1} (b). 
The last term, $- d(\ve n \wedge \ve \dot{v})\cdot \ve e_z$, is torque due to acceleration of the co-moving frame.

\begin{table*}[]
  \footnotesize
  \caption{Dimensional and non-dimensional parameters. The length scale of the glider is denoted by $a$, and the kinematic fluid viscosity by $\nu$. Gliding ant in air: parameters from \citet{yanoviak2005Directed}. Ocean glider prototype~\cite{Mitchell2013}.
  We take $a$ to be the half length of its cylindrical body. The aspect ratio is defined as the ratio between the length and the diameter of the cylindrical body. The density is determined based on the reported mass and the volume of the cylinder body. Glider in silicon oil (this paper) with mass density slightly larger than that of the fluid.} 
  \label{tab:parameters}
  \begin{tabular}{lccccccc}\hline\hline
    Case                                           & $a~[\si{cm}]$ & $\nu~[\si{m^2/s}]$        & \makecell{Glider density \\ $[\si{kg/m^3}]$} & \makecell{Fluid density\\$[\si{kg/m^3}]$} & $\lambda$ & $\beta$     & Ga         \\ \hline
    Gliding ant in air~\cite{yanoviak2005Directed} & 0.6           & \num{1.8E-05}           & 290                           & 1.2                           & 2         & 240       & 1200       \\
    Ocean glider prototype~\cite{Mitchell2013}     & 25             & \num{1.0E-06}            & 950 to 1500                    & 1000                          & 5          & 0.95 to 1.5 & 0 to $2.8\times10^4$ \\
    Glider in silicon oil                          & 5             & \num{2e-3} to \num{1E-04} & 950 to 1500                    & 764                           & $> 1$         & 1.2 to 2.0  & 8 to 340   \\
    \hline\hline
  \end{tabular}
\end{table*}

The dynamics in Eq.~(\ref{eq:eom}) additionally depends on the aspect ratio $\lambda$ and mass-density ratio $\beta$, both introduced above, as well as the Galileo number $\Ga=\sqrt{(\beta-1)ga^3}/\nu$, which characterises the fluid-solid interactions governing $\ve F_{\rm h}$ and $\ve T_{\rm h}$.
These three parameters are equivalent to the non-dimensional parameters used in  Ref.~\cite{bhowmick2024inertia} to describe the motion for a spheroid settling in a quiescent fluid.
We parameterise the dynamics using $\Ga$ since it is uniquely determined by intrinsic particle and flow properties, while $\Rep$, depending on the actual settling speed, must be evaluated {\it a posteriori}.
For small $\Ga$, $\Rep$ scales as $\Ga^2$, but this relation overestimates $\Rep$ for larger $\Ga$.
Table~\ref{tab:parameters} presents example parameters~\cite{yanoviak2005Directed,Mitchell2013} and a suggestion for future glider experiments.
Table~\ref{tab:parameters} shows that $\Ga$ varies over several orders of magnitude, but only navigation strategies for $10^3$ were addressed using the large-$\Ga$ empirical model in earlier studies~\cite{Paoletti2011,Novati2019}. 
Our fully resolved simulations enable exploration of small and moderate $\Ga$, while high $\Ga$ remain too computationally costly. We therefore focus on this unexplored regime, using $\Ga= 4.4, 35.4, 283$, which can be realized experimentally in viscous liquids such as silicon oil.
The corresponding values of the particle Reynolds number are $\Rep\approx 3, 31$, and $240$.
We use $\beta=2$ in our simulations to consider settling gliders. The numerical method used in the DNS restricts the possible aspect ratios we can study. Here we report results for $\lambda=2$.

\subsection{Numerical method}\label{section:dns}
To solve Eq.~(\ref{eq:eom}), we calculate the hydrodynamic force $\boldsymbol{F}_{\rm h}$ and torque $T_{\rm h}$ using DNS that fully resolve particle-fluid interactions via the immersed boundary method~\cite{Peskin2002}. The fluid velocity $\ve u$ and pressure $p$ are obtained by solving the incompressible Navier-Stokes equations on an Eulerian grid (in dimensional units):
\begin{equation} \label{eq2.1}
  \begin{array}{c}
    \nabla\cdot{\ve u} = 0\,, \\
    \rho_{\rm f}\left(\frac{\partial\ve u}{\partial t} + \ve u\cdot\ve\nabla\ve u\right)=-\ve\nabla p + \mu \nabla ^2\ve u+\rho_{\rm f}{\ve f}\,.
  \end{array}
\end{equation}
Here $\mu$ is the dynamic viscosity of the fluid, and the immersed-boundary forcing term $\ve f$ accounts for the disturbance caused by the motion of the glider.
It is obtained by enforcing no-slip condition at the fluid-solid interface using the direct-forcing immersed boundary method~\cite{Uhlmann2005,Breugem2012}. 
In this approach, fluid velocity is interpolated from the Eulerian grid to uniformly distributed Lagrangian marker points on the glider surface. Immersed boundary forces are computed from the difference between interpolated and rigid-body velocities, then spread back onto the Eulerian grid to yield $\ve f$ in Eq.~(\ref{eq2.1}). See~\citet{Breugem2012} for details.
In this framework, $\boldsymbol{F}_{\rm h}$ and $\boldsymbol{T}_{\rm h}$ are calculated as~\cite{Breugem2012}:
\begin{equation} \label{eq2.3}
  {{\boldsymbol{F}}_{\rm{h}}} =  - {\rho _f}\int_{{{\mathscr V}_{\rm{p}}}} {\boldsymbol{f}} {\rm{d}}V + {\rho _f}\frac{{\rm{d}}}{{{\rm{d}}t}}\int_{{{\mathscr V}_{\rm{p}}}} {\boldsymbol{u}} {\rm{d}}V \,,
\end{equation}
\begin{equation} \label{eq2.4}
  {{\boldsymbol{T}}_{\rm h}} =  - {\rho _f}\int_{ {\mathscr{V}_{\rm p}}} {\boldsymbol{r}_c\wedge\boldsymbol{f}}{\rm d}V+{\rho _f}\frac{\mathrm{d} }{{\mathrm{d} t}}\int_{\mathscr{V}_{\rm p}} {{\boldsymbol{r}_c\wedge\boldsymbol{u}}}{\rm d}V \,.
\end{equation}
Here, $\mathscr{V}_{\rm p}$ is the volume of the glider and $\boldsymbol{r}_c$ is the vector from the geometric centre to the point of integration.

We solve Eqs.~(\ref{eq2.1}) using a second-order central difference method~\cite{Kim2002} on a uniform Eulerian grid of size $24a\times32a$. The grid resolution $\Delta x$ depends on the Galileo number: $\Delta x=a/16$ for $\Ga=4.4$ and 35.4, and $\Delta x=a/32$ for $\Ga=283$. The time step size is $\Delta t=0.014 \sqrt{a/g}$.
Simulations are converged with respect to domain size, grid spacing and time step.
Periodic boundary conditions are applied horizontally. Vertically, we impose Neumann conditions at the top and Dirichlet conditions at the bottom to enforce zero fluid velocity, modeling quiescent fluid.
This bottom condition does not represent a solid wall because the domain follows the glider downwards, preventing it to reach the lower boundary.
See Ref.~\cite{jiang2024flow} for further details on the numerical method.

\subsection{Reinforcement learning}
We use double deep Q-learning~\cite{Mnih2015,Hasselt2016,Sutton2018,mehlig2021machine} to find optimal control strategies for navigating a glider.
Unlike previous study~\cite{Novati2019} that target a fixed position, we train a single strategy for reaching targets at different positions, demonstrating robustness.

As illustrated in Fig.~\ref{fig1} (c), the glider is released from rest at $\ve r_0 = (x_0, y_0)$ with random orientation and is tasked to reach a target at $\ve r_{\rm T} = (x_{\rm T}, y_{\rm T})$, where $y_{\rm T} - y_0 = -40$ and $x_{\rm T} - x_0$ is drawn uniformly from $[-30, 30]$. At regular intervals $\tau_{\rm u} = 1.41$, the glider selects one of five center-of-mass equilibrium positions ($d_{\rm e} = 0, \pm0.2, \pm0.4$) based on its relative position to the target, velocity, tilt angle, and angular velocity. The update interval $\tau_{\rm u}$ is long enough for the mass to reach its new position but short enough to allow over 30 decisions per trial.

To optimise navigation, we choose the reward:
\begin{equation} \label{eq4.2}
  r =  - {h_1}\frac{\left( {{x} - {x_{\rm T}}} \right)^2}{a^2}{\left( {\frac{y-y_0}{y_{\rm T}-y_0}} \right)^2} + {h_2}\Theta(y_{\rm T} - y){e^{ - \left| {{x} - {x_{\rm T}}} \right|/(2a)}}\,,
\end{equation}
with coefficients $h_1=0.125$ and $h_2=100$. The first term in Eq.~(\ref{eq4.2}) penalises the horizontal distance to the target, scaled by $[(y-y_0)/(y_{\rm T}-y_0)]^2$ to reduce the penalty when the glider is far above the target, promoting exploration.
The second term is a one-time bonus based on horizontal distance to the target when the glider reaches the target height ($y=y_{\rm T}$), at which point the episode terminates. Here $\Theta(\cdot)$ is the Heaviside function.

Training details and hyperparameters are provided in Appendix~\ref{section:rl}.

\section{Results}
Figure~\ref{fig2} documents the success of reinforcement learning for three values of $\Ga$. Landing positions $x$ under the learned strategy are shown for 200 different targets $x_{\rm T}$. For the smallest Ga, only targets within a narrow range are reachable [Fig.~\ref{fig2} (a)].
For the intermediate value of Ga, the glider can reach any target
within the entire range of $\left|x_{\rm T}-x_0\right|$ [Fig.~\ref{fig2} (b)]. For the largest Ga, the range narrows again [Fig.~\ref{fig2} (c)].
\begin{figure}
  \includegraphics[width=0.5\textwidth]{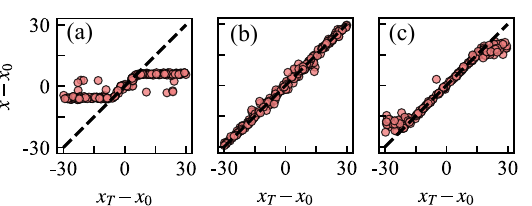}
  \caption{\label{fig2} Landing position $x$ versus randomly prescribed target points $x_{\rm T}$. The dashed line represents perfect navigation $x=x_{\rm T}$. Parameters: (a) $\Ga=4.4$; (b) $\Ga=35.4$, and (c) $\Ga=283$.
  }
\end{figure}

\begin{figure}
  \includegraphics[width=0.45\textwidth]{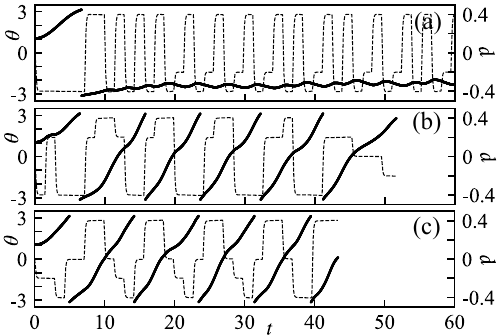}
  \caption{Glider orientation for optimal strategies at different $\Ga$. The control $d(t)$ (dashed lines) and the tilt angle $\theta(t)$ (solid lines) are shown against time. (a) $\Ga=4.4$; (b) $\Ga=35.4$; (c) $\Ga=283$.}
  \label{fig:traj}
\end{figure}

\begin{figure*}
  \includegraphics[width=0.9\textwidth]{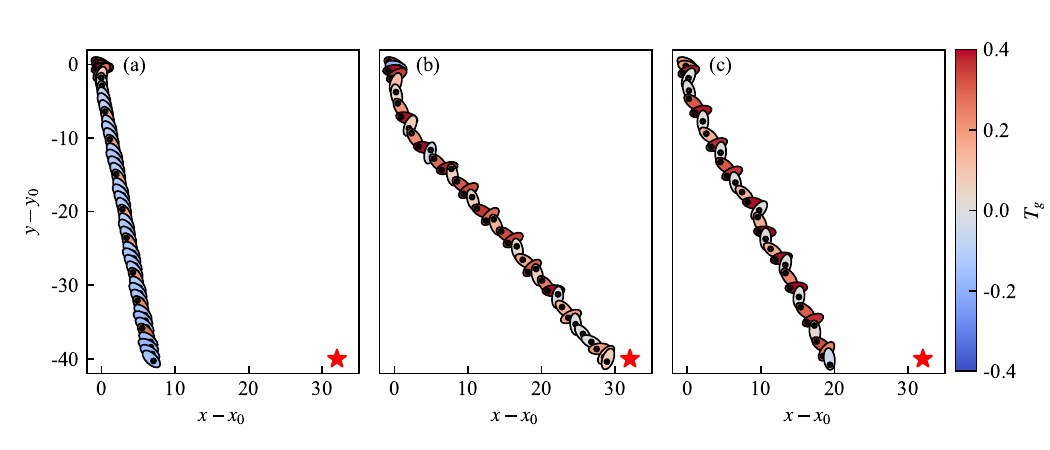}
  \caption{Glider trajectories corresponding to Fig.~\ref{fig:traj} (a--c). Ellipses outline the glider at time intervals of action update, $\tau_{\rm u} = 1.41$, coloured by instantaneous gravity torque $T_g$. Black dots mark the centre-of-mass, and the red star indicates the target. The glider size is enlarged by a factor 1.5 for better visualisation. (a) $\Ga=4.4$; (b) $\Ga=35.4$; (c) $\Ga=283$.}
  \label{fig:traj-pos}
\end{figure*}

The optimal strategies differ for different Ga. This becomes clear from Fig.~\ref{fig:traj}, which shows optimal strategies for a far target at $x_{\rm T}=32$.
At the lowest value of $\Ga$, the glider
learns to maintain an inclined orientation by adjusting its centre-of-mass in a periodic fashion [Fig.~\ref{fig:traj} (a)], so that it glides at an approximately constant angle. At larger values of $\Ga$, the glider learns to move its centre-of-mass to tumble anticlockwise (clockwise) with an approximately constant angular velocity to reach a target to its right [Fig.~\ref{fig:traj} (b,c)] (left, respectively, not shown).
The corresponding trajectories are shown in Fig.~\ref{fig:traj-pos}.
In short, by adjusting its centre-of-mass, the glider manages to navigate by two different strategies, either inclined settling or tumbling.

The same two strategies were found using an empirical model designed for gliders settling at large particle Reynolds numbers~\cite{Paoletti2011,Novati2019}, of the order of $10^3$. \citet{Novati2019} found that tumbling outperforms inclined settling at these Reynolds numbers for small enough aspect ratio $\lambda$, or large enough  density ratio~$\beta$. \citet{Paoletti2011} used a model similar to Eq.~(\ref{eq::torquemodel}) for the torque and a corresponding model for the force acting on the glider.  There are three main differences to these works. First, DNS allows us to investigate how the optimal strategy changes as the particle Reynolds number changes (at fixed $\lambda$ and $\beta$). Second, we use an explicit form of control (the moving centre-of-mass), instead of an arbitrary control torque. 
Third, we consider a glider in a highly viscous fluid (not heavy gliders settling in air). The reason is that laboratory experiments are likely more feasible for more viscous fluids, simply because the glider tends to settle more slowly.

In order to understand how different strategies emerge at different
Reynolds numbers, we analyse the torques in Eq.~(\ref{eq::dynr}). In this equation both $T_{\rm g}$ and the last term depend on the control through $d\sim d_{\rm e}$. However, the last term is small because $\ve v$ approaches the steady settling velocity after
an initial transient. The hydrodynamical torque $T_{\rm h}$ is modeled using an empirical torque model, Eq.~(1) in Ref.~\cite{ern2012WakeInduceda}:
\begin{equation}
  T_{\rm h} = -c_1 \omega -c_2 |\omega|\omega - c_3 |v|^2 \sin\phi \cos\phi - c_4 \dot{\omega}\,.
  \label{eq::torquemodel}
\end{equation}
We note that this model applies to both two- and three-dimensional objects which have three mutually orthogonal symmetry planes and move in quiescent fluids~\cite{Andersen2005,ern2012WakeInduceda}, although the coefficients are quantitatively different.
The first term in Eq.~(\ref{eq::torquemodel}) is Stokes torque~\cite{happel1983Low}, the second term describes a second-order correction to the Stokes torque~\cite{pierson2021Hydrodynamic}, similar to the Oseen correction to the Stokes drag.
The first two terms together represent the so-called dissipative torque~\cite{Andersen2005,pesavento2004Falling}.
The third term describes a fluid-inertia torque,  consistent with the form obtained in small $\Rep$-perturbation theory~\cite{khayat1989Inertia,lopez2017Inertial,gustavsson2021Effect,pierson2021Hydrodynamic,bhowmick2024inertia}, where $\phi$ is the angle between the orientation $\ve n$ and velocity $\ve v$ of the glider.
The angle $\phi$ differs from the tilt angle  $\theta$ in Eq.~(\ref{eq::dynr}). We note that Refs.~\cite{bhowmick2024inertia,gustavsson2021Effect} considered the case where $\phi\approx\theta$.
The fourth term is added moment of inertia resulting from acceleration of the surrounding fluid~\cite{lamb1924Hydrodynamics}.
Related models were first used for passively settling particles~\cite{pesavento2004Falling,Andersen2005,pesavento2004Falling}, with coefficients chosen to generate qualitatively accurate trajectories at $\Rep \sim 10^3$. 
\begin{table}[t]
  \caption{\label{tab::coeff} Numerical values of the coefficients in Eq.~(\ref{eq::torquemodel}), obtained by fitting the DNS torque  to Eq.~(\ref{eq::torquemodel}) for $d=0$.
  }
  \begin{tabular}{lcccc}\hline\hline
    $ \Ga     $ & $c_1$ & $c_2$ & $c_3$ & $c_4$ \\
    \hline
    $4.4$       & 0.192 & 0.911 & 0.151 & 0.070 \\
    $35.4$      & 0.024 & 0.203 & 0.143 & 0.070 \\
    $283$       & 0.066 & 0.183 & 0.096 & 0.070 \\
    \hline\hline
  \end{tabular}
\end{table}

As explained above, we use the Galileo number to parameterise particle-fluid interactions in our model, instead of the Re$_p$. The two numbers are in a one-to-one correspondence, but Ga has the advantage that it is 
uniquely determined by intrinsic particle and flow properties, while $\Rep$ must be evaluated \emph{a posteriori}.
To study how $T_{\rm h}$ depends on $\Ga$, we fix the coefficient $c_4$ in Eq.~(\ref{eq::torquemodel}) to the known value for an elliptical plate in the potential flow limit~\cite{Sedov1980}, 
\begin{equation}
    c_4 = \tfrac{1}{16\beta} \lambda(1-\lambda^{-2})^2\,.
    \label{eq:c4}
\end{equation} 
The remaining coefficients $c_1$, $c_2$, and $c_3$ are fitted for each $\Ga$ to two sets of DNS trajectories.
The first set consists of trajectories with $d=0$ and different initial conditions, typically resulting in small but non-zero angular velocities.
The second set is a single tumbling trajectory following the heuristic strategy Eq.~(\ref{eq:tumbling}) (explained later).
We weigh the two sets by their number of trajectories during fitting, yielding accurate torque estimates for both gliding and tumbling (see Appendix~\ref{sec:empirical_model}, Figure~\ref{fig::fitting}).

The resulting coefficients are shown in Table~\ref{tab::coeff}.
\citet{Andersen2005} suggest that the coefficients $c_1$ and $c_2$  decrease as Ga increases. Our results in Table~\ref{tab::coeff} are broadly consistent
with that expectation, except $c_1$ which increases slightly from Ga$=35.4$ to $283$.
Table~\ref{tab::coeff}  shows that $c_3$ decreases as Ga increases. Our values of $c_3$ are quantitatively consistent with the coefficients reported by~\cite{oh2022Drag} which relies on DNS of rotating elliptical cylinders in an uniform flow, with errors less than~$15\%$.

At our largest Galileo number, $\Ga=283$ ($\Rep \approx 240$), the glider learns to tumble, just  like the gliders from Refs.~\cite{Paoletti2011,Novati2019}, with $\Rep\sim 10^3$.
At high Re$_{\rm p}$ the rapidly tumbling glider generates a lift force, similar to the Magnus effect observed for a rotating sphere~\cite{rubinow1961Transverse}, where rotation breaks the flow symmetry and produces a lift force $\ve{F}_{\rm L} \sim \ve{\omega}\wedge \ve{v}$. Here, the lift force on the glider provides horizontal momentum for navigation, and a vertical component that balances part of the gravity force, slowing down the settling.
Because the direction of the lift force depends on the direction of angular velocity, the glider learns to tumble anticlockwise to reach any target to its right, and tumble clockwise for the target to its left.
Our DNS shows that the lift force becomes larger if the glider rotates with larger $|\omega|$, similar to a sphere~\cite{rubinow1961Transverse}. Therefore, it is important to understand how the gravitational (control) torque $T_g$ and the hydrodynamical torque $T_{\rm h}$ from Eq.~(\ref{eq::torquemodel}) interact in Eq.~(\ref{eq::dynr}) to generate large $|\ve\omega|$.
\citet{Paoletti2011} identified a mechanism for this using a model with a constant control torque, where tumbling emerges from  balancing this torque with the average dissipative torque. In our case, by contrast, the control torque changes as a function of time, simply
because the tilt angle and the centre-of-mass displacement change with time.
How does the glider learn to tumble under these circumstances?  Fig.~\ref{fig:traj} (c) shows that
the glider shifts its centre-of-mass when the angle $\theta$ reaches $\theta = 0$ or $\pi$, where the sign of $T_g=-d\sin\theta$ is about to change.
This strategy generates a  gravity torque that is mostly positive [Fig.~\ref{fig:traj-pos} (a)], a prerequisite for persistent anticlockwise tumbling.

\begin{figure}
\includegraphics[width=0.45\textwidth]{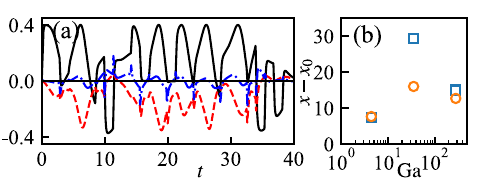}
\caption{\label{fig::dist-theta} (a) Torques for $\Ga=283$, for the trajectory shown in Fig.~\ref{fig:traj-pos} (c). Solid line: $T_g$; dashed line: dissipative torque $-c_1 \omega -c_2 |\omega|\omega$; dash-dotted line: fluid-inertia torque, $-c_3 |v|^2 \sin\phi \cos\phi$. The added moment of inertia torque, $-c_4\dot{\omega}$, is negligible (not shown). (b) Maximal horizontal distance covered by a glider when it uses the tumbling strategy [Eq.~(\ref{eq:tumbling})] ($\square$), or settling with a certain fixed orientation ($\circ $).}
\end{figure}

How can this mechanism be understood from Eq.~(\ref{eq::torquemodel})? For the tumbling strategy, Fig.~\ref{fig::dist-theta} (a) shows
the two largest contributions to the hydrodynamic torque $T_{\rm h}$: the dissipative and the fluid-inertia torques. We see that the latter is significantly smaller than the former, although $|\ve v|$ and $|\ve\omega|$ are of  the same order for the simulated
trajectories (not shown). The difference in magnitude is explained by the fact that $\sin\phi \cos\phi$ fluctuates around
zero, and that the prefactor $c_3$ is smaller than $c_2$ by a factor of two (Table~\ref{tab::coeff}). In conclusion, the fluid-inertia torque has only a minor effect on the dynamics.
Fig.~\ref{fig::dist-theta} (a) indicates that  the control torque $T_g$ is approximately balanced by the dissipative torque on average (the same conclusion holds for intermediate Galileo number, $\Ga=35.4$).
From this balance, we obtain an estimate of $\omega$ which differs by 20\% to 30\% from our DNS results.
This shows that the model is broadly consistent with our DNS, even though the control torque depends on time, unlike Ref.~\cite{Paoletti2011} which assumed constant control.

At small Ga,  the glider navigates by maintaining an approximately constant angle $\theta$, resulting in a horizontal force which allows steering.
The horizontal force originates from the anisotropic drag coefficient of an elliptical glider. For a glider with $\lambda=2$ at $\Ga=4.4$, the maximal drag coefficient (achieved when $\ve n\perp\ve v$, or equivalently when $\phi=\pi$) is only about $30\%$ larger than the minimal drag coefficient (achieved when $\ve n\parallel\ve v$, or $\phi = 0$), which constrains the horizontal travel distance even if the glider settles with an optimal orientation.
For a constant control $T_g$~\cite{Paoletti2011}, settling with optimal orientation can be achieved by balancing the fluid-inertia torque \cite{khayat1989Inertia,lopez2017Inertial,gustavsson2021Effect,pierson2021Hydrodynamic,bhowmick2024inertia}, the third term on the r.h.s of Eq.~(\ref{eq::torquemodel}),  against $T_g$~\cite{Roy2019}.
However, this balance may be hard to achieve because it requires very small yet precise values of $T_g$ at small Ga. In practice it is easier to have a control torque $T_g$ that assumes only discrete values, as in our model. Our DNS for $\Ga=4.4$ show that $T_g$ is one order of magnitude larger than fluid-inertia torque for the smallest non-zero offset ($d=0.2$).
Fig.~\ref{fig:traj} (a) shows that the glider learns to adjust its centre-of-mass dynamically to stabilise its tilt angle, and we see that the fluctuations around the average are small. This average angle is the optimum that maximises the horizontal traveling distance, verified by measuring the distance in DNS of a glider that settles with fixed $\theta$ at different values of $\theta$.

Now we discuss how the reachable horizontal range depends on Ga (Fig.~\ref{fig2}).
To this end, we consider a heuristic tumbling strategy,
\begin{equation}
  d(t) = {\rm sgn}[\theta(t) ]d_{\rm max}\,,
  \label{eq:tumbling}
\end{equation}
which allows the glider to maximise its rotation speed by changing its centre-of-mass whenever the sign of $\theta$ changes, inspired by the strategy learned by reinforcement learning shown in Fig.~\ref{fig2} (b,c).
Fig.~\ref{fig::dist-theta} (b) shows DNS results for the reachable range for the heuristic tumbling strategy~(\ref{eq:tumbling}) as a function of Ga ($\square$).
Also shown are DNS results for the reachable range obtained by gliding at a fixed optimal angle $\theta$ which maximises the horizontal distance travelled by the glider.
At large Ga, tumbling is better than gliding at fixed $\theta$, because the glider can achieve large angular velocities and  the dissipative torque is smaller compared to the case of smaller Ga  [Eq.~(\ref{eq::torquemodel}) and Table~\ref{tab::coeff}].  This explains why the glider learns to tumble
at large Ga.
But why does the reachable range for the tumbling strategy decrease as $\Ga$ increases from 35.4 to 283? Our DNS show that the glider rotates at almost the same speed at both values of $\Ga$ because of similar dissipative torque coefficients, but the lift force at $\Ga=283$ is $40\%$ smaller than at $\Ga=35.4$. We verified by DNS that the main reason for the decreased lift force is a decrease in the pressure contribution to this force.
Finally, at the smallest Ga, gliding and tumbling have approximately equal reachable ranges. Nevertheless, our reinforcement-learning runs show that the gliding strategy always wins at small Ga (even when gliding is slightly suboptimal).
Tumbling is hard to learn, because the glider must manage to rotate in the correct direction, and to stop tumbling when it approaches the target.

Here, we consider a two-dimensional glider. We found this necessary because our direct numerical simulations combined with reinforcement learning are too computationally expensive in three spatial dimensions. We expect that the navigation strategies remain valid for a three-dimensional glider, because the empirical torque model~(\ref{eq::torquemodel}) extends to a three-dimensional compact particle~\cite{Andersen2005,ern2012WakeInduceda}. Since the coefficients are different in three dimensions, the crossover between the two strategies (inclined settling versus tumbling) may occur at a different Reynolds number, compared with two dimensions.

How can our results be checked by laboratory experiments?
For a glider of size $a=5~\si{cm}$ -- large enough to carry  sensors, motors, and batteries~\cite{jaffe2017swarm} -- our range of $\Ga$ can be realised using silicone oil with viscosity ranging from $2\times 10^{-3}~\si{m^2s^{-1}}$ to $1\times 10^{-4}~\si{m^2s^{-1}}$ (Table~\ref{tab:parameters}). This yields a settling speed of about $1~\si{m/s}$, angular velocity $10~\si{rad/s}$, and total fall height $2~\si{m}$.
We expect that small changes in $\beta$ (from 1.2 to 2, see Table~\ref{tab:parameters}) does not make a qualitative difference to the settling dynamics, because the main effect of changing  $\beta$ is on the buoyancy force $\ve{F}_b$. Decreasing $\beta$ increases buoyancy. This reduces the settling speed of the glider, leading to longer traveling times. Moreover, although the added mass torque is also proportional to $\beta^{-1}$ [Eq.~(\ref{eq:c4})], its magnitude smaller than that  of the other torques. So the fact that the added-mass torque changes as $\beta$  changes is less important.

Here, we assume an infinitely large domain. How hydrodynamic forces and torques change near a wall or the floor of the container can be addressed by implementing the correct boundary conditions in the simulations.
Moreover, we assume that the glider senses its position relative to the target, its orientation, velocity, and angular velocity. For practical applications it is beneficial to reduce the number of signals. We expect that $x(t)-x_{\rm T}(t)$, $\theta(t)$, and $\omega(t)$ are the most important signals (Appendix~\ref{section:signals}), because $x$ indicates the horizontal direction to travel, while $\theta$ and $\omega$ are needed to control the rotational dynamics.
These signals can be measured with a small motion processing unit~\cite{invensense2013mpu6050}.

Controlling an autonomous glider by its moving centre-of-mass has distinct advantages over other control methods. First, it generates a control torque without energy-demanding propulsion. Moving the centre-of-mass within a small particle only requires a small amount of energy. Second, the glider exploits gravity to generate its control torque, which requires no other external field to be applied on the glider for navigation, such as commonly used magnetic or electric fields~\cite{jiang2022Controla}. A drawback is that the magnitude of the gravity torque is constrained by the geometry of the glider and its instantaneous orientation, which impede the maximal travel distance. This limitation may be eased by adding shape control of the glider, which has been shown to increase stability in natural gliders such as birds~\cite{harvey2019Wing}.

\section{Conclusions}
In this paper, we investigated how a compact glider  navigates with an adjustable centre-of-mass as it settles in a viscous fluid at different Reynolds numbers. Using direct numerical simulations, we found that the glider learns two distinct strategies to move large
horizontal distances: it either settles with an inclined orientation optimized for horizontal displacement, or it tumbles. Which of the two strategies is optimal depends sensitively
on $\Rep$ for given shape and mass density of the glider. 
For small $\Rep$, inclined settling is better, while the glider learns to tumble at larger $\Rep$, generating a horizontal lift force that  extends its travel range.

Both strategies were previously found using an empirical model designed for gliders settling at large particle Reynolds numbers~\cite{Paoletti2011,Novati2019}. Besides the fact that our direct numerical simulations allowed us to study how the optimal strategy depends on the particle Reynolds number, there are two more important differences to Refs.~\cite{Paoletti2011,Novati2019}. First, we used an explicit control torque (the moving centre-of-mass) rather than an arbitrary one, and we could show that how the glider learns to execute these strategies by adjusting its centre-of-mass. It settles with an inclined orientation by moving the centre-of-mass back and forth, or tumbles by adjusting its centre-of-mass as a function of its orientation [Eq.~(\ref{eq:tumbling})].
It was not {\em a priori} clear that this works, because the amplitude of the explicit control is constrained
by the maximal distance the centre-of-mass can move, and the control mechanism has a delay. Nevertheless,
the glider can land precisely on any target within its maximal horizontal travel range by first approaching the target using the appropriate strategy, and then relying on viscous drag to slow down once it is above the target. Second, with laboratory experiments in mind, and because large aspect ratios $\lambda$ are hard to simulate, we considered a compact glider in a viscous fluid with mass-density ratio $\beta$ of order unity, rather than a glider with 
large $\lambda$ and $\beta$.

Like earlier studies~\cite{Paoletti2011,Novati2019}, we considered a two-dimensional elliptical glider. 
In our case, this is necessary because direct numerical simulations in three dimensions in combination with the reinforcement-learning algorithm were too computationally expensive. We expect that the tumbling and gliding mechanisms survive for a three-dimensional glider, because the empirical torque model (\ref{eq::torquemodel}) extends to three-dimensional particles~\cite{Andersen2005,ern2012WakeInduceda}, albeit with different coefficients.  We cannot exclude that the extra degrees of freedom in three dimensions may allow for additional possibly more complicated strategies.

In nature, settling particles with more complicated shapes are common~\cite{candelier2025torques}, even chiral~\cite{Kim2021,huseby2025helical}. 
Such particles experience translation-rotation coupling even in the Stokes limit, causing the  glider to rotate  as it settles. This could allow new strategies with increased navigation range, as well as additional control possibilities. Moreover, one should also consider shape changes
as a control mechanism, because this changes the hydrodynamic torques \cite{Candelier2016,Roy2019,roy2023orientation,ravichandran2023orientation,maches2024settling,candelier2025torques}
the glider can use for control. Examples in nature are gliding ants that change posture~\cite{yanoviak2005Directed}, and birds that control the morphology of their wings~\cite{harvey2019Wing}.

\begin{acknowledgments}
We acknowledge support from Vetenskapsr\aa{}det,  grant nos. 2018-03974, 2023-03617 (JQ and KG), and 2021-4452 (BM). KG, BM and JR acknowledge support from the Knut and Alice Wallenberg Foundation, grant no. 2019.0079. XJ and LZ  were supported by the Natural Science Foundation of China through grants 92252104 and 12388101. Our collaboration was supported by the joint China-Sweden mobility programme [National Natural Science Foundation of China (NSFC)-Swedish Foundation for International Cooperation in Research and Higher Education (STINT)] through grant nos. 11911530141 (NSFC) and CH2018-7737 (STINT).
\end{acknowledgments}

\bibliography{ref}

\providecommand{\noopsort}[1]{}\providecommand{\singleletter}[1]{#1}%
\begin{thebibliography}{49}%
\makeatletter
\providecommand \@ifxundefined [1]{%
 \@ifx{#1\undefined}
}%
\providecommand \@ifnum [1]{%
 \ifnum #1\expandafter \@firstoftwo
 \else \expandafter \@secondoftwo
 \fi
}%
\providecommand \@ifx [1]{%
 \ifx #1\expandafter \@firstoftwo
 \else \expandafter \@secondoftwo
 \fi
}%
\providecommand \natexlab [1]{#1}%
\providecommand \enquote  [1]{``#1''}%
\providecommand \bibnamefont  [1]{#1}%
\providecommand \bibfnamefont [1]{#1}%
\providecommand \citenamefont [1]{#1}%
\providecommand \href@noop [0]{\@secondoftwo}%
\providecommand \href [0]{\begingroup \@sanitize@url \@href}%
\providecommand \@href[1]{\@@startlink{#1}\@@href}%
\providecommand \@@href[1]{\endgroup#1\@@endlink}%
\providecommand \@sanitize@url [0]{\catcode `\\12\catcode `\$12\catcode
  `\&12\catcode `\#12\catcode `\^12\catcode `\_12\catcode `\%12\relax}%
\providecommand \@@startlink[1]{}%
\providecommand \@@endlink[0]{}%
\providecommand \url  [0]{\begingroup\@sanitize@url \@url }%
\providecommand \@url [1]{\endgroup\@href {#1}{\urlprefix }}%
\providecommand \urlprefix  [0]{URL }%
\providecommand \Eprint [0]{\href }%
\providecommand \doibase [0]{http://dx.doi.org/}%
\providecommand \selectlanguage [0]{\@gobble}%
\providecommand \bibinfo  [0]{\@secondoftwo}%
\providecommand \bibfield  [0]{\@secondoftwo}%
\providecommand \translation [1]{[#1]}%
\providecommand \BibitemOpen [0]{}%
\providecommand \bibitemStop [0]{}%
\providecommand \bibitemNoStop [0]{.\EOS\space}%
\providecommand \EOS [0]{\spacefactor3000\relax}%
\providecommand \BibitemShut  [1]{\csname bibitem#1\endcsname}%
\let\auto@bib@innerbib\@empty
\bibitem [{\citenamefont {Yanoviak}\ \emph {et~al.}(2005)\citenamefont
  {Yanoviak}, \citenamefont {Dudley},\ and\ \citenamefont
  {Kaspari}}]{yanoviak2005Directed}%
  \BibitemOpen
  \bibfield  {author} {\bibinfo {author} {\bibfnamefont {S.~P.}\ \bibnamefont
  {Yanoviak}}, \bibinfo {author} {\bibfnamefont {R.}~\bibnamefont {Dudley}}, \
  and\ \bibinfo {author} {\bibfnamefont {M.}~\bibnamefont {Kaspari}},\
  }\bibfield  {title} {\enquote {\bibinfo {title} {Directed aerial descent in
  canopy ants},}\ }\href@noop {} {\bibfield  {journal} {\bibinfo  {journal}
  {Nature}\ }\textbf {\bibinfo {volume} {433}},\ \bibinfo {pages} {624--626}
  (\bibinfo {year} {2005})}\BibitemShut {NoStop}%
\bibitem [{\citenamefont {Cummins}\ \emph {et~al.}(2018)\citenamefont
  {Cummins}, \citenamefont {Seale}, \citenamefont {Macente}, \citenamefont
  {Certini}, \citenamefont {Mastropaolo}, \citenamefont {Viola},\ and\
  \citenamefont {Nakayama}}]{cummins2018Separated}%
  \BibitemOpen
  \bibfield  {author} {\bibinfo {author} {\bibfnamefont {C.}~\bibnamefont
  {Cummins}}, \bibinfo {author} {\bibfnamefont {M.}~\bibnamefont {Seale}},
  \bibinfo {author} {\bibfnamefont {A.}~\bibnamefont {Macente}}, \bibinfo
  {author} {\bibfnamefont {D.}~\bibnamefont {Certini}}, \bibinfo {author}
  {\bibfnamefont {E.}~\bibnamefont {Mastropaolo}}, \bibinfo {author}
  {\bibfnamefont {I.~M.}\ \bibnamefont {Viola}}, \ and\ \bibinfo {author}
  {\bibfnamefont {N.}~\bibnamefont {Nakayama}},\ }\bibfield  {title} {\enquote
  {\bibinfo {title} {A separated vortex ring underlies the flight of the
  dandelion},}\ }\href@noop {} {\bibfield  {journal} {\bibinfo  {journal}
  {Nature}\ }\textbf {\bibinfo {volume} {562}},\ \bibinfo {pages} {414--418}
  (\bibinfo {year} {2018})}\BibitemShut {NoStop}%
\bibitem [{\citenamefont {Augspurger}(1986)}]{augspurger1986Morphology}%
  \BibitemOpen
  \bibfield  {author} {\bibinfo {author} {\bibfnamefont {C.~K.}\ \bibnamefont
  {Augspurger}},\ }\bibfield  {title} {\enquote {\bibinfo {title} {Morphology
  and dispersal potential of wind-dispersed diaspores of neotropical trees},}\
  }\href@noop {} {\bibfield  {journal} {\bibinfo  {journal} {American Journal
  of Botany}\ }\textbf {\bibinfo {volume} {73}},\ \bibinfo {pages} {353--363}
  (\bibinfo {year} {1986})}\BibitemShut {NoStop}%
\bibitem [{\citenamefont {Kim}\ \emph {et~al.}(2021)\citenamefont {Kim},
  \citenamefont {Li}, \citenamefont {Kim}, \citenamefont {Park}, \citenamefont
  {Jang} \emph {et~al.}}]{Kim2021}%
  \BibitemOpen
  \bibfield  {author} {\bibinfo {author} {\bibfnamefont {B.~H.}\ \bibnamefont
  {Kim}}, \bibinfo {author} {\bibfnamefont {K.}~\bibnamefont {Li}}, \bibinfo
  {author} {\bibfnamefont {J.}~\bibnamefont {Kim}}, \bibinfo {author}
  {\bibfnamefont {Y.}~\bibnamefont {Park}}, \bibinfo {author} {\bibfnamefont
  {H.}~\bibnamefont {Jang}},  \emph {et~al.},\ }\bibfield  {title} {\enquote
  {\bibinfo {title} {Three-dimensional electronic microfliers inspired by
  wind-dispersed seeds},}\ }\href@noop {} {\bibfield  {journal} {\bibinfo
  {journal} {Nature}\ }\textbf {\bibinfo {volume} {597}},\ \bibinfo {pages}
  {503--510} (\bibinfo {year} {2021})}\BibitemShut {NoStop}%
\bibitem [{\citenamefont {Kahn}\ \emph {et~al.}(1999)\citenamefont {Kahn},
  \citenamefont {Katz},\ and\ \citenamefont {Pister}}]{kahn1999Next}%
  \BibitemOpen
  \bibfield  {author} {\bibinfo {author} {\bibfnamefont {J.~M.}\ \bibnamefont
  {Kahn}}, \bibinfo {author} {\bibfnamefont {R.~H.}\ \bibnamefont {Katz}}, \
  and\ \bibinfo {author} {\bibfnamefont {K.~S.~J.}\ \bibnamefont {Pister}},\
  }\bibfield  {title} {\enquote {\bibinfo {title} {Next century challenges:
  {{Mobile}} networking for smart dust},}\ }in\ \href@noop {} {\emph {\bibinfo
  {booktitle} {Proceedings of the 5th Annual {{ACM}}/{{IEEE}} International
  Conference on {{Mobile}} Computing and Networking}}}\ (\bibinfo  {publisher}
  {ACM},\ \bibinfo {address} {Seattle Washington USA},\ \bibinfo {year}
  {1999})\ pp.\ \bibinfo {pages} {271--278}\BibitemShut {NoStop}%
\bibitem [{\citenamefont {Lermusiaux}\ \emph {et~al.}(2017)\citenamefont
  {Lermusiaux}, \citenamefont {Subramani}, \citenamefont {Lin}, \citenamefont
  {Kulkarni}, \citenamefont {Gupta}, \citenamefont {Dutt}, \citenamefont
  {Lolla}, \citenamefont {Haley}, \citenamefont {Ali}, \citenamefont
  {Mirabito},\ and\ \citenamefont {Jana}}]{Lermusiaux2017}%
  \BibitemOpen
  \bibfield  {author} {\bibinfo {author} {\bibfnamefont {P.~F.~J.}\
  \bibnamefont {Lermusiaux}}, \bibinfo {author} {\bibfnamefont {D.~N.}\
  \bibnamefont {Subramani}}, \bibinfo {author} {\bibfnamefont {J.}~\bibnamefont
  {Lin}}, \bibinfo {author} {\bibfnamefont {C.~S.}\ \bibnamefont {Kulkarni}},
  \bibinfo {author} {\bibfnamefont {A.}~\bibnamefont {Gupta}}, \bibinfo
  {author} {\bibfnamefont {A.}~\bibnamefont {Dutt}}, \bibinfo {author}
  {\bibfnamefont {T.}~\bibnamefont {Lolla}}, \bibinfo {author} {\bibfnamefont
  {{\relax Jr}.}~\bibnamefont {Haley}, \bibfnamefont {P.~J.}}, \bibinfo
  {author} {\bibfnamefont {W.~H.}\ \bibnamefont {Ali}}, \bibinfo {author}
  {\bibfnamefont {C.}~\bibnamefont {Mirabito}}, \ and\ \bibinfo {author}
  {\bibfnamefont {S.}~\bibnamefont {Jana}},\ }\bibfield  {title} {\enquote
  {\bibinfo {title} {A future for intelligent autonomous ocean observing
  systems},}\ }\href@noop {} {\bibfield  {journal} {\bibinfo  {journal}
  {Journal of Marine Research}\ }\textbf {\bibinfo {volume} {75}},\ \bibinfo
  {pages} {765--813} (\bibinfo {year} {2017})}\BibitemShut {NoStop}%
\bibitem [{\citenamefont {Leonard}\ \emph {et~al.}(2007)\citenamefont
  {Leonard}, \citenamefont {Paley}, \citenamefont {Lekien}, \citenamefont
  {Sepulchre}, \citenamefont {Fratantoni},\ and\ \citenamefont
  {Davis}}]{leonard2007collective}%
  \BibitemOpen
  \bibfield  {author} {\bibinfo {author} {\bibfnamefont {Naomi~Ehrich}\
  \bibnamefont {Leonard}}, \bibinfo {author} {\bibfnamefont {Derek~A}\
  \bibnamefont {Paley}}, \bibinfo {author} {\bibfnamefont {Francois}\
  \bibnamefont {Lekien}}, \bibinfo {author} {\bibfnamefont {Rodolphe}\
  \bibnamefont {Sepulchre}}, \bibinfo {author} {\bibfnamefont {David~M}\
  \bibnamefont {Fratantoni}}, \ and\ \bibinfo {author} {\bibfnamefont {Russ~E}\
  \bibnamefont {Davis}},\ }\bibfield  {title} {\enquote {\bibinfo {title}
  {Collective motion, sensor networks, and ocean sampling},}\ }\href@noop {}
  {\bibfield  {journal} {\bibinfo  {journal} {Proceedings of the IEEE}\
  }\textbf {\bibinfo {volume} {95}},\ \bibinfo {pages} {48--74} (\bibinfo
  {year} {2007})}\BibitemShut {NoStop}%
\bibitem [{\citenamefont {Mitchell}\ \emph {et~al.}(2013)\citenamefont
  {Mitchell}, \citenamefont {Wilkening},\ and\ \citenamefont
  {Mahmoudian}}]{Mitchell2013}%
  \BibitemOpen
  \bibfield  {author} {\bibinfo {author} {\bibfnamefont {B.}~\bibnamefont
  {Mitchell}}, \bibinfo {author} {\bibfnamefont {E.}~\bibnamefont {Wilkening}},
  \ and\ \bibinfo {author} {\bibfnamefont {N.}~\bibnamefont {Mahmoudian}},\
  }\bibfield  {title} {\enquote {\bibinfo {title} {Low cost underwater gliders
  for littoral marine research},}\ }in\ \href@noop {} {\emph {\bibinfo
  {booktitle} {American Control Conference ({{ACC}})}}},\ \bibinfo {series and
  number} {Proceedings of the American Control Conference}\ (\bibinfo {year}
  {2013})\ pp.\ \bibinfo {pages} {1412--1417}\BibitemShut {NoStop}%
\bibitem [{\citenamefont {Happel}\ and\ \citenamefont
  {Brenner}(1983)}]{happel1983Low}%
  \BibitemOpen
  \bibfield  {author} {\bibinfo {author} {\bibfnamefont {J.}~\bibnamefont
  {Happel}}\ and\ \bibinfo {author} {\bibfnamefont {H.}~\bibnamefont
  {Brenner}},\ }\href@noop {} {\emph {\bibinfo {title} {Low {{Reynolds}} Number
  Hydrodynamics: {{With}} Special Applications to Particulate Media}}}\
  (\bibinfo  {publisher} {Springer Science \& Business Media},\ \bibinfo
  {address} {Berlin, Germany},\ \bibinfo {year} {1983})\BibitemShut {NoStop}%
\bibitem [{\citenamefont {Pierson}\ \emph {et~al.}(2021)\citenamefont
  {Pierson}, \citenamefont {Kharrouba},\ and\ \citenamefont
  {Magnaudet}}]{pierson2021Hydrodynamic}%
  \BibitemOpen
  \bibfield  {author} {\bibinfo {author} {\bibfnamefont {J.}~\bibnamefont
  {Pierson}}, \bibinfo {author} {\bibfnamefont {M.}~\bibnamefont {Kharrouba}},
  \ and\ \bibinfo {author} {\bibfnamefont {J.}~\bibnamefont {Magnaudet}},\
  }\bibfield  {title} {\enquote {\bibinfo {title} {Hydrodynamic torque on a
  slender cylinder rotating perpendicularly to its symmetry axis},}\
  }\href@noop {} {\bibfield  {journal} {\bibinfo  {journal} {Physical Review
  Fluids}\ }\textbf {\bibinfo {volume} {6}},\ \bibinfo {pages} {094303}
  (\bibinfo {year} {2021})}\BibitemShut {NoStop}%
\bibitem [{\citenamefont {Khayat}\ and\ \citenamefont
  {Cox}(1989)}]{khayat1989Inertia}%
  \BibitemOpen
  \bibfield  {author} {\bibinfo {author} {\bibfnamefont {R.~E.}\ \bibnamefont
  {Khayat}}\ and\ \bibinfo {author} {\bibfnamefont {R.~G.}\ \bibnamefont
  {Cox}},\ }\bibfield  {title} {\enquote {\bibinfo {title} {Inertia effects on
  the motion of long slender bodies},}\ }\href@noop {} {\bibfield  {journal}
  {\bibinfo  {journal} {Journal of Fluid Mechanics}\ }\textbf {\bibinfo
  {volume} {209}},\ \bibinfo {pages} {435--462} (\bibinfo {year}
  {1989})}\BibitemShut {NoStop}%
\bibitem [{\citenamefont {Lopez}\ and\ \citenamefont
  {Guazzelli}(2017)}]{lopez2017Inertial}%
  \BibitemOpen
  \bibfield  {author} {\bibinfo {author} {\bibfnamefont {D.}~\bibnamefont
  {Lopez}}\ and\ \bibinfo {author} {\bibfnamefont {E.}~\bibnamefont
  {Guazzelli}},\ }\bibfield  {title} {\enquote {\bibinfo {title} {Inertial
  effects on fibers settling in a vortical flow},}\ }\href@noop {} {\bibfield
  {journal} {\bibinfo  {journal} {Physical Review Fluids}\ }\textbf {\bibinfo
  {volume} {2}},\ \bibinfo {pages} {024306} (\bibinfo {year}
  {2017})}\BibitemShut {NoStop}%
\bibitem [{\citenamefont {Gustavsson}\ \emph {et~al.}(2021)\citenamefont
  {Gustavsson}, \citenamefont {Sheikh}, \citenamefont {Naso}, \citenamefont
  {Pumir},\ and\ \citenamefont {Mehlig}}]{gustavsson2021Effect}%
  \BibitemOpen
  \bibfield  {author} {\bibinfo {author} {\bibfnamefont {K.}~\bibnamefont
  {Gustavsson}}, \bibinfo {author} {\bibfnamefont {M.~Z.}\ \bibnamefont
  {Sheikh}}, \bibinfo {author} {\bibfnamefont {A.}~\bibnamefont {Naso}},
  \bibinfo {author} {\bibfnamefont {A.}~\bibnamefont {Pumir}}, \ and\ \bibinfo
  {author} {\bibfnamefont {B.}~\bibnamefont {Mehlig}},\ }\bibfield  {title}
  {\enquote {\bibinfo {title} {Effect of particle inertia on the alignment of
  small ice crystals in turbulent clouds},}\ }\href@noop {} {\bibfield
  {journal} {\bibinfo  {journal} {Journal of the Atmospheric Sciences}\
  }\textbf {\bibinfo {volume} {78}},\ \bibinfo {pages} {2573--2587} (\bibinfo
  {year} {2021})}\BibitemShut {NoStop}%
\bibitem [{\citenamefont {Candelier}\ \emph {et~al.}(2023)\citenamefont
  {Candelier}, \citenamefont {Mehaddi}, \citenamefont {Mehlig},\ and\
  \citenamefont {Magnaudet}}]{candelier2023second}%
  \BibitemOpen
  \bibfield  {author} {\bibinfo {author} {\bibfnamefont {F.}~\bibnamefont
  {Candelier}}, \bibinfo {author} {\bibfnamefont {R.}~\bibnamefont {Mehaddi}},
  \bibinfo {author} {\bibfnamefont {B.}~\bibnamefont {Mehlig}}, \ and\ \bibinfo
  {author} {\bibfnamefont {J.}~\bibnamefont {Magnaudet}},\ }\bibfield  {title}
  {\enquote {\bibinfo {title} {Second-order inertial forces and torques on a
  sphere in a viscous steady linear flow},}\ }\href@noop {} {\bibfield
  {journal} {\bibinfo  {journal} {Journal of Fluid Mechanics}\ }\textbf
  {\bibinfo {volume} {954}},\ \bibinfo {pages} {A25} (\bibinfo {year}
  {2023})}\BibitemShut {NoStop}%
\bibitem [{\citenamefont {Ern}\ \emph {et~al.}(2012)\citenamefont {Ern},
  \citenamefont {Risso}, \citenamefont {Fabre},\ and\ \citenamefont
  {Magnaudet}}]{ern2012WakeInduceda}%
  \BibitemOpen
  \bibfield  {author} {\bibinfo {author} {\bibfnamefont {P.}~\bibnamefont
  {Ern}}, \bibinfo {author} {\bibfnamefont {F.}~\bibnamefont {Risso}}, \bibinfo
  {author} {\bibfnamefont {D.}~\bibnamefont {Fabre}}, \ and\ \bibinfo {author}
  {\bibfnamefont {J.}~\bibnamefont {Magnaudet}},\ }\bibfield  {title} {\enquote
  {\bibinfo {title} {Wake-induced oscillatory paths of bodies freely rising or
  falling in fluids},}\ }\href@noop {} {\bibfield  {journal} {\bibinfo
  {journal} {Annual Review of Fluid Mechanics}\ }\textbf {\bibinfo {volume}
  {44}},\ \bibinfo {pages} {97--121} (\bibinfo {year} {2012})}\BibitemShut
  {NoStop}%
\bibitem [{\citenamefont {Pesavento}\ and\ \citenamefont
  {Wang}(2004)}]{pesavento2004Falling}%
  \BibitemOpen
  \bibfield  {author} {\bibinfo {author} {\bibfnamefont {U.}~\bibnamefont
  {Pesavento}}\ and\ \bibinfo {author} {\bibfnamefont {Z.~J.}\ \bibnamefont
  {Wang}},\ }\bibfield  {title} {\enquote {\bibinfo {title} {Falling paper
  {{Navier-Stokes}} solutions model of fluid forces and center of mass
  elevation},}\ }\href@noop {} {\bibfield  {journal} {\bibinfo  {journal}
  {Physical Review Letters}\ }\textbf {\bibinfo {volume} {93}},\ \bibinfo
  {pages} {144501} (\bibinfo {year} {2004})}\BibitemShut {NoStop}%
\bibitem [{\citenamefont {Andersen}\ \emph {et~al.}(2005)\citenamefont
  {Andersen}, \citenamefont {Pesavento},\ and\ \citenamefont
  {Wang}}]{Andersen2005}%
  \BibitemOpen
  \bibfield  {author} {\bibinfo {author} {\bibfnamefont {A.}~\bibnamefont
  {Andersen}}, \bibinfo {author} {\bibfnamefont {U.}~\bibnamefont {Pesavento}},
  \ and\ \bibinfo {author} {\bibfnamefont {Z.~J.}\ \bibnamefont {Wang}},\
  }\bibfield  {title} {\enquote {\bibinfo {title} {Analysis of transitions
  between fluttering, tumbling and steady descent of falling cards},}\
  }\href@noop {} {\bibfield  {journal} {\bibinfo  {journal} {Journal of Fluid
  Mechanics}\ }\textbf {\bibinfo {volume} {541}},\ \bibinfo {pages} {91--104}
  (\bibinfo {year} {2005})}\BibitemShut {NoStop}%
\bibitem [{\citenamefont {Field}\ \emph {et~al.}(1997)\citenamefont {Field},
  \citenamefont {Klaus}, \citenamefont {Moore},\ and\ \citenamefont
  {Nori}}]{field1997Chaotic}%
  \BibitemOpen
  \bibfield  {author} {\bibinfo {author} {\bibfnamefont {S.~B.}\ \bibnamefont
  {Field}}, \bibinfo {author} {\bibfnamefont {M.}~\bibnamefont {Klaus}},
  \bibinfo {author} {\bibfnamefont {M.~G.}\ \bibnamefont {Moore}}, \ and\
  \bibinfo {author} {\bibfnamefont {F.}~\bibnamefont {Nori}},\ }\bibfield
  {title} {\enquote {\bibinfo {title} {Chaotic dynamics of falling disks},}\
  }\href@noop {} {\bibfield  {journal} {\bibinfo  {journal} {Nature}\ }\textbf
  {\bibinfo {volume} {388}},\ \bibinfo {pages} {252--254} (\bibinfo {year}
  {1997})}\BibitemShut {NoStop}%
\bibitem [{\citenamefont {Paoletti}\ and\ \citenamefont
  {Mahadevan}(2011)}]{Paoletti2011}%
  \BibitemOpen
  \bibfield  {author} {\bibinfo {author} {\bibfnamefont {P.}~\bibnamefont
  {Paoletti}}\ and\ \bibinfo {author} {\bibfnamefont {L.}~\bibnamefont
  {Mahadevan}},\ }\bibfield  {title} {\enquote {\bibinfo {title} {Planar
  controlled gliding, tumbling and descent},}\ }\href@noop {} {\bibfield
  {journal} {\bibinfo  {journal} {Journal of Fluid Mechanics}\ }\textbf
  {\bibinfo {volume} {689}},\ \bibinfo {pages} {489--516} (\bibinfo {year}
  {2011})}\BibitemShut {NoStop}%
\bibitem [{\citenamefont {Mnih}\ \emph {et~al.}(2015)\citenamefont {Mnih},
  \citenamefont {Kavukcuoglu}, \citenamefont {Silver}, \citenamefont {Rusu},
  \citenamefont {Veness}, \citenamefont {Bellemare}, \citenamefont {Graves},
  \citenamefont {Riedmiller}, \citenamefont {Fidjeland}, \citenamefont
  {Ostrovski}, \citenamefont {Petersen}, \citenamefont {Beattie}, \citenamefont
  {Sadik}, \citenamefont {Antonoglou}, \citenamefont {King}, \citenamefont
  {Kumaran}, \citenamefont {Wierstra}, \citenamefont {Legg},\ and\
  \citenamefont {Hassabis}}]{Mnih2015}%
  \BibitemOpen
  \bibfield  {author} {\bibinfo {author} {\bibfnamefont {V.}~\bibnamefont
  {Mnih}}, \bibinfo {author} {\bibfnamefont {K.}~\bibnamefont {Kavukcuoglu}},
  \bibinfo {author} {\bibfnamefont {D.}~\bibnamefont {Silver}}, \bibinfo
  {author} {\bibfnamefont {Andrei~A.}\ \bibnamefont {Rusu}}, \bibinfo {author}
  {\bibfnamefont {J.}~\bibnamefont {Veness}}, \bibinfo {author} {\bibfnamefont
  {M.~G.}\ \bibnamefont {Bellemare}}, \bibinfo {author} {\bibfnamefont
  {A.}~\bibnamefont {Graves}}, \bibinfo {author} {\bibfnamefont
  {M.}~\bibnamefont {Riedmiller}}, \bibinfo {author} {\bibfnamefont {A.~K.}\
  \bibnamefont {Fidjeland}}, \bibinfo {author} {\bibfnamefont {G.}~\bibnamefont
  {Ostrovski}}, \bibinfo {author} {\bibfnamefont {S.}~\bibnamefont {Petersen}},
  \bibinfo {author} {\bibfnamefont {C.}~\bibnamefont {Beattie}}, \bibinfo
  {author} {\bibfnamefont {A.}~\bibnamefont {Sadik}}, \bibinfo {author}
  {\bibfnamefont {I.}~\bibnamefont {Antonoglou}}, \bibinfo {author}
  {\bibfnamefont {H.}~\bibnamefont {King}}, \bibinfo {author} {\bibfnamefont
  {D.}~\bibnamefont {Kumaran}}, \bibinfo {author} {\bibfnamefont
  {D.}~\bibnamefont {Wierstra}}, \bibinfo {author} {\bibfnamefont
  {S.}~\bibnamefont {Legg}}, \ and\ \bibinfo {author} {\bibfnamefont
  {D.}~\bibnamefont {Hassabis}},\ }\bibfield  {title} {\enquote {\bibinfo
  {title} {Human-level control through deep reinforcement learning},}\
  }\href@noop {} {\bibfield  {journal} {\bibinfo  {journal} {Nature}\ }\textbf
  {\bibinfo {volume} {518}},\ \bibinfo {pages} {529--533} (\bibinfo {year}
  {2015})}\BibitemShut {NoStop}%
\bibitem [{\citenamefont {Mehlig}(2021)}]{mehlig2021machine}%
  \BibitemOpen
  \bibfield  {author} {\bibinfo {author} {\bibfnamefont {B.}~\bibnamefont
  {Mehlig}},\ }\href@noop {} {\emph {\bibinfo {title} {Machine Learning with
  Neural Networks: {{An}} Introduction for Scientists and Engineers}}}\
  (\bibinfo  {publisher} {Cambridge University Press},\ \bibinfo {address}
  {Cambridge},\ \bibinfo {year} {2021})\BibitemShut {NoStop}%
\bibitem [{\citenamefont {Colabrese}\ \emph {et~al.}(2017)\citenamefont
  {Colabrese}, \citenamefont {Gustavsson}, \citenamefont {Celani},\ and\
  \citenamefont {Biferale}}]{Colabrese2017}%
  \BibitemOpen
  \bibfield  {author} {\bibinfo {author} {\bibfnamefont {S.}~\bibnamefont
  {Colabrese}}, \bibinfo {author} {\bibfnamefont {K.}~\bibnamefont
  {Gustavsson}}, \bibinfo {author} {\bibfnamefont {A.}~\bibnamefont {Celani}},
  \ and\ \bibinfo {author} {\bibfnamefont {L.}~\bibnamefont {Biferale}},\
  }\bibfield  {title} {\enquote {\bibinfo {title} {Flow navigation by smart
  microswimmers via reinforcement learning},}\ }\href@noop {} {\bibfield
  {journal} {\bibinfo  {journal} {Physical Review Letters}\ }\textbf {\bibinfo
  {volume} {128}},\ \bibinfo {pages} {158004} (\bibinfo {year}
  {2017})}\BibitemShut {NoStop}%
\bibitem [{\citenamefont {Novati}\ \emph {et~al.}(2019)\citenamefont {Novati},
  \citenamefont {Mahadevan},\ and\ \citenamefont {Koumoutsakos}}]{Novati2019}%
  \BibitemOpen
  \bibfield  {author} {\bibinfo {author} {\bibfnamefont {G.}~\bibnamefont
  {Novati}}, \bibinfo {author} {\bibfnamefont {L.}~\bibnamefont {Mahadevan}}, \
  and\ \bibinfo {author} {\bibfnamefont {P.}~\bibnamefont {Koumoutsakos}},\
  }\bibfield  {title} {\enquote {\bibinfo {title} {Controlled gliding and
  perching through deep-reinforcement-learning},}\ }\href@noop {} {\bibfield
  {journal} {\bibinfo  {journal} {Physical Review Fluids}\ }\textbf {\bibinfo
  {volume} {4}},\ \bibinfo {pages} {093902} (\bibinfo {year}
  {2019})}\BibitemShut {NoStop}%
\bibitem [{\citenamefont {Gunnarson}\ \emph {et~al.}(2021)\citenamefont
  {Gunnarson}, \citenamefont {Mandralis}, \citenamefont {Novati}, \citenamefont
  {Koumoutsakos},\ and\ \citenamefont {Dabiri}}]{Gunnarson2021}%
  \BibitemOpen
  \bibfield  {author} {\bibinfo {author} {\bibfnamefont {P.}~\bibnamefont
  {Gunnarson}}, \bibinfo {author} {\bibfnamefont {I.}~\bibnamefont
  {Mandralis}}, \bibinfo {author} {\bibfnamefont {G.}~\bibnamefont {Novati}},
  \bibinfo {author} {\bibfnamefont {P.}~\bibnamefont {Koumoutsakos}}, \ and\
  \bibinfo {author} {\bibfnamefont {J.~O.}\ \bibnamefont {Dabiri}},\ }\bibfield
   {title} {\enquote {\bibinfo {title} {Learning efficient navigation in
  vortical flow fields},}\ }\href@noop {} {\bibfield  {journal} {\bibinfo
  {journal} {Nature Communications}\ }\textbf {\bibinfo {volume} {12}}
  (\bibinfo {year} {2021})}\BibitemShut {NoStop}%
\bibitem [{\citenamefont {Alageshan}\ \emph {et~al.}(2020)\citenamefont
  {Alageshan}, \citenamefont {Verma}, \citenamefont {Bec},\ and\ \citenamefont
  {Pandit}}]{Alageshan2020}%
  \BibitemOpen
  \bibfield  {author} {\bibinfo {author} {\bibfnamefont {J.~K.}\ \bibnamefont
  {Alageshan}}, \bibinfo {author} {\bibfnamefont {A.~K.}\ \bibnamefont
  {Verma}}, \bibinfo {author} {\bibfnamefont {J.}~\bibnamefont {Bec}}, \ and\
  \bibinfo {author} {\bibfnamefont {R.}~\bibnamefont {Pandit}},\ }\bibfield
  {title} {\enquote {\bibinfo {title} {Machine learning strategies for
  path-planning microswimmers in turbulent flows},}\ }\href@noop {} {\bibfield
  {journal} {\bibinfo  {journal} {Physical Review E}\ }\textbf {\bibinfo
  {volume} {101}} (\bibinfo {year} {2020})}\BibitemShut {NoStop}%
\bibitem [{\citenamefont {Jiang}\ \emph {et~al.}(2022)\citenamefont {Jiang},
  \citenamefont {Yang}, \citenamefont {Ferreira},\ and\ \citenamefont
  {Zhang}}]{jiang2022Controla}%
  \BibitemOpen
  \bibfield  {author} {\bibinfo {author} {\bibfnamefont {J.}~\bibnamefont
  {Jiang}}, \bibinfo {author} {\bibfnamefont {Z.}~\bibnamefont {Yang}},
  \bibinfo {author} {\bibfnamefont {A.}~\bibnamefont {Ferreira}}, \ and\
  \bibinfo {author} {\bibfnamefont {L.}~\bibnamefont {Zhang}},\ }\bibfield
  {title} {\enquote {\bibinfo {title} {Control and {{Autonomy}} of
  {{Microrobots}}: {{Recent Progress}} and {{Perspective}}},}\ }\href@noop {}
  {\bibfield  {journal} {\bibinfo  {journal} {Advanced Intelligent Systems}\
  }\textbf {\bibinfo {volume} {4}},\ \bibinfo {pages} {2100279} (\bibinfo
  {year} {2022})}\BibitemShut {NoStop}%
\bibitem [{\citenamefont {Roy}\ \emph {et~al.}(2019)\citenamefont {Roy},
  \citenamefont {Hamati}, \citenamefont {Tierney}, \citenamefont {Koch},\ and\
  \citenamefont {Voth}}]{Roy2019}%
  \BibitemOpen
  \bibfield  {author} {\bibinfo {author} {\bibfnamefont {A.}~\bibnamefont
  {Roy}}, \bibinfo {author} {\bibfnamefont {R.~J.}\ \bibnamefont {Hamati}},
  \bibinfo {author} {\bibfnamefont {L.}~\bibnamefont {Tierney}}, \bibinfo
  {author} {\bibfnamefont {D.~L.}\ \bibnamefont {Koch}}, \ and\ \bibinfo
  {author} {\bibfnamefont {G.~A.}\ \bibnamefont {Voth}},\ }\bibfield  {title}
  {\enquote {\bibinfo {title} {Inertial torques and a symmetry breaking
  orientational transition in the sedimentation of slender fibres},}\
  }\href@noop {} {\bibfield  {journal} {\bibinfo  {journal} {Journal of Fluid
  Mechanics}\ }\textbf {\bibinfo {volume} {875}},\ \bibinfo {pages} {576--596}
  (\bibinfo {year} {2019})}\BibitemShut {NoStop}%
\bibitem [{\citenamefont {Jiang}\ \emph
  {et~al.}(2024{\natexlab{a}})\citenamefont {Jiang}, \citenamefont {Xu},\ and\
  \citenamefont {Zhao}}]{jiangSettling2024}%
  \BibitemOpen
  \bibfield  {author} {\bibinfo {author} {\bibfnamefont {X.}~\bibnamefont
  {Jiang}}, \bibinfo {author} {\bibfnamefont {C.}~\bibnamefont {Xu}}, \ and\
  \bibinfo {author} {\bibfnamefont {L.}~\bibnamefont {Zhao}},\ }\bibfield
  {title} {\enquote {\bibinfo {title} {Settling and collision of spheroidal
  particles with an offset mass centre in a quiescent fluid},}\ }\href@noop {}
  {\bibfield  {journal} {\bibinfo  {journal} {Journal of Fluid Mechanics}\
  }\textbf {\bibinfo {volume} {984}},\ \bibinfo {pages} {A40} (\bibinfo {year}
  {2024}{\natexlab{a}})}\BibitemShut {NoStop}%
\bibitem [{\citenamefont {Bhowmick}\ \emph {et~al.}(2024)\citenamefont
  {Bhowmick}, \citenamefont {Seesing}, \citenamefont {Gustavsson},
  \citenamefont {Guettler}, \citenamefont {Wang}, \citenamefont {Pumir},
  \citenamefont {Mehlig},\ and\ \citenamefont {Bagheri}}]{bhowmick2024inertia}%
  \BibitemOpen
  \bibfield  {author} {\bibinfo {author} {\bibfnamefont {T.}~\bibnamefont
  {Bhowmick}}, \bibinfo {author} {\bibfnamefont {J.}~\bibnamefont {Seesing}},
  \bibinfo {author} {\bibfnamefont {K.}~\bibnamefont {Gustavsson}}, \bibinfo
  {author} {\bibfnamefont {J.}~\bibnamefont {Guettler}}, \bibinfo {author}
  {\bibfnamefont {Y.}~\bibnamefont {Wang}}, \bibinfo {author} {\bibfnamefont
  {A.}~\bibnamefont {Pumir}}, \bibinfo {author} {\bibfnamefont
  {B.}~\bibnamefont {Mehlig}}, \ and\ \bibinfo {author} {\bibfnamefont
  {G.}~\bibnamefont {Bagheri}},\ }\bibfield  {title} {\enquote {\bibinfo
  {title} {Inertia induces strong orientation fluctuations of nonspherical
  atmospheric particles},}\ }\href@noop {} {\bibfield  {journal} {\bibinfo
  {journal} {Physical Review Letters}\ }\textbf {\bibinfo {volume} {132}},\
  \bibinfo {pages} {034101} (\bibinfo {year} {2024})}\BibitemShut {NoStop}%
\bibitem [{\citenamefont {Peskin}(2002)}]{Peskin2002}%
  \BibitemOpen
  \bibfield  {author} {\bibinfo {author} {\bibfnamefont {C.~S.}\ \bibnamefont
  {Peskin}},\ }\bibfield  {title} {\enquote {\bibinfo {title} {The immersed
  boundary method},}\ }\href@noop {} {\bibfield  {journal} {\bibinfo  {journal}
  {Acta Numerica}\ }\textbf {\bibinfo {volume} {11}},\ \bibinfo {pages}
  {479--517} (\bibinfo {year} {2002})}\BibitemShut {NoStop}%
\bibitem [{\citenamefont {Uhlmann}(2005)}]{Uhlmann2005}%
  \BibitemOpen
  \bibfield  {author} {\bibinfo {author} {\bibfnamefont {M.}~\bibnamefont
  {Uhlmann}},\ }\bibfield  {title} {\enquote {\bibinfo {title} {An immersed
  boundary method with direct forcing for the simulation of particulate
  flows},}\ }\href@noop {} {\bibfield  {journal} {\bibinfo  {journal} {Journal
  of Computational Physics}\ }\textbf {\bibinfo {volume} {209}},\ \bibinfo
  {pages} {448--476} (\bibinfo {year} {2005})}\BibitemShut {NoStop}%
\bibitem [{\citenamefont {Breugem}(2012)}]{Breugem2012}%
  \BibitemOpen
  \bibfield  {author} {\bibinfo {author} {\bibfnamefont {W.}~\bibnamefont
  {Breugem}},\ }\bibfield  {title} {\enquote {\bibinfo {title} {A second-order
  accurate immersed boundary method for fully resolved simulations of
  particle-laden flows},}\ }\href@noop {} {\bibfield  {journal} {\bibinfo
  {journal} {Journal of Computational Physics}\ }\textbf {\bibinfo {volume}
  {231}},\ \bibinfo {pages} {4469--4498} (\bibinfo {year} {2012})}\BibitemShut
  {NoStop}%
\bibitem [{\citenamefont {Kim}\ \emph {et~al.}(2002)\citenamefont {Kim},
  \citenamefont {Baek},\ and\ \citenamefont {Sung}}]{Kim2002}%
  \BibitemOpen
  \bibfield  {author} {\bibinfo {author} {\bibfnamefont {K.}~\bibnamefont
  {Kim}}, \bibinfo {author} {\bibfnamefont {S.~J.}\ \bibnamefont {Baek}}, \
  and\ \bibinfo {author} {\bibfnamefont {H.~J.}\ \bibnamefont {Sung}},\
  }\bibfield  {title} {\enquote {\bibinfo {title} {An implicit velocity
  decoupling procedure for the incompressible {Navier-Stokes} equations},}\
  }\href@noop {} {\bibfield  {journal} {\bibinfo  {journal} {International
  Journal for Numerical Methods in Fluids}\ }\textbf {\bibinfo {volume} {38}},\
  \bibinfo {pages} {125--138} (\bibinfo {year} {2002})}\BibitemShut {NoStop}%
\bibitem [{\citenamefont {Jiang}\ \emph
  {et~al.}(2024{\natexlab{b}})\citenamefont {Jiang}, \citenamefont {Huang},
  \citenamefont {Xu},\ and\ \citenamefont {Zhao}}]{jiang2024flow}%
  \BibitemOpen
  \bibfield  {author} {\bibinfo {author} {\bibfnamefont {X.}~\bibnamefont
  {Jiang}}, \bibinfo {author} {\bibfnamefont {W.}~\bibnamefont {Huang}},
  \bibinfo {author} {\bibfnamefont {C.}~\bibnamefont {Xu}}, \ and\ \bibinfo
  {author} {\bibfnamefont {L.}~\bibnamefont {Zhao}},\ }\bibfield  {title}
  {\enquote {\bibinfo {title} {A flow-reconstruction based approach for the
  computation of hydrodynamic stresses on immersed body surface},}\ }\href@noop
  {} {\bibfield  {journal} {\bibinfo  {journal} {Journal of Computational
  Physics}\ }\textbf {\bibinfo {volume} {508}},\ \bibinfo {pages} {113025}
  (\bibinfo {year} {2024}{\natexlab{b}})}\BibitemShut {NoStop}%
\bibitem [{\citenamefont {van Hasselt}\ \emph {et~al.}(2016)\citenamefont {van
  Hasselt}, \citenamefont {Guez},\ and\ \citenamefont {Silver}}]{Hasselt2016}%
  \BibitemOpen
  \bibfield  {author} {\bibinfo {author} {\bibfnamefont {H.}~\bibnamefont {van
  Hasselt}}, \bibinfo {author} {\bibfnamefont {A.}~\bibnamefont {Guez}}, \ and\
  \bibinfo {author} {\bibfnamefont {D.}~\bibnamefont {Silver}},\ }\bibfield
  {title} {\enquote {\bibinfo {title} {Deep reinforcement learning with double
  q-learning},}\ }\href@noop {} {\bibfield  {journal} {\bibinfo  {journal}
  {Proceedings of the AAAI Conference on Artificial Intelligence}\ }\textbf
  {\bibinfo {volume} {30}} (\bibinfo {year} {2016})}\BibitemShut {NoStop}%
\bibitem [{\citenamefont {Sutton}\ and\ \citenamefont
  {Barto}(2018)}]{Sutton2018}%
  \BibitemOpen
  \bibfield  {author} {\bibinfo {author} {\bibfnamefont {R.~S.}\ \bibnamefont
  {Sutton}}\ and\ \bibinfo {author} {\bibfnamefont {A.~G.}\ \bibnamefont
  {Barto}},\ }\href@noop {} {\emph {\bibinfo {title} {Reinforcement Learning:
  {{An}} Introduction}}},\ \bibinfo {edition} {2nd}\ ed.\ (\bibinfo
  {publisher} {The MIT Press},\ \bibinfo {year} {2018})\BibitemShut {NoStop}%
\bibitem [{\citenamefont {Lamb}(1924)}]{lamb1924Hydrodynamics}%
  \BibitemOpen
  \bibfield  {author} {\bibinfo {author} {\bibfnamefont {H.}~\bibnamefont
  {Lamb}},\ }\href@noop {} {\emph {\bibinfo {title} {Hydrodynamics}}}\
  (\bibinfo  {publisher} {University Press},\ \bibinfo {year}
  {1924})\BibitemShut {NoStop}%
\bibitem [{\citenamefont {Sedov}(1980)}]{Sedov1980}%
  \BibitemOpen
  \bibfield  {author} {\bibinfo {author} {\bibfnamefont {L.~I.}\ \bibnamefont
  {Sedov}},\ }\href@noop {} {\emph {\bibinfo {title} {Two-Dimensional Problems
  of Hydrodynamics and Aerodynamics}}}\ (\bibinfo  {publisher} {Moscow Izdatel
  Nauka},\ \bibinfo {year} {1980})\BibitemShut {NoStop}%
\bibitem [{\citenamefont {Oh}\ \emph {et~al.}(2022)\citenamefont {Oh},
  \citenamefont {Park},\ and\ \citenamefont {Choi}}]{oh2022Drag}%
  \BibitemOpen
  \bibfield  {author} {\bibinfo {author} {\bibfnamefont {G.}~\bibnamefont
  {Oh}}, \bibinfo {author} {\bibfnamefont {H.}~\bibnamefont {Park}}, \ and\
  \bibinfo {author} {\bibfnamefont {J.}~\bibnamefont {Choi}},\ }\bibfield
  {title} {\enquote {\bibinfo {title} {Drag, lift, and torque coefficients for
  various geometrical configurations of elliptic cylinder under {{Stokes}} to
  laminar flow regimes},}\ }\href@noop {} {\bibfield  {journal} {\bibinfo
  {journal} {AIP Advances}\ }\textbf {\bibinfo {volume} {12}},\ \bibinfo
  {pages} {065228} (\bibinfo {year} {2022})}\BibitemShut {NoStop}%
\bibitem [{\citenamefont {Rubinow}\ and\ \citenamefont
  {Keller}(1961)}]{rubinow1961Transverse}%
  \BibitemOpen
  \bibfield  {author} {\bibinfo {author} {\bibfnamefont {S.~I.}\ \bibnamefont
  {Rubinow}}\ and\ \bibinfo {author} {\bibfnamefont {Joseph~B.}\ \bibnamefont
  {Keller}},\ }\bibfield  {title} {\enquote {\bibinfo {title} {The transverse
  force on a spinning sphere moving in a viscous fluid},}\ }\href@noop {}
  {\bibfield  {journal} {\bibinfo  {journal} {Journal of Fluid Mechanics}\
  }\textbf {\bibinfo {volume} {11}},\ \bibinfo {pages} {447--459} (\bibinfo
  {year} {1961})}\BibitemShut {NoStop}%
\bibitem [{\citenamefont {Jaffe}\ \emph {et~al.}(2017)\citenamefont {Jaffe},
  \citenamefont {Franks}, \citenamefont {Roberts}, \citenamefont {Mirza},
  \citenamefont {Schurgers}, \citenamefont {Kastner},\ and\ \citenamefont
  {Boch}}]{jaffe2017swarm}%
  \BibitemOpen
  \bibfield  {author} {\bibinfo {author} {\bibfnamefont {J.~S}\ \bibnamefont
  {Jaffe}}, \bibinfo {author} {\bibfnamefont {P.~JS}\ \bibnamefont {Franks}},
  \bibinfo {author} {\bibfnamefont {P.~LD}\ \bibnamefont {Roberts}}, \bibinfo
  {author} {\bibfnamefont {D.}~\bibnamefont {Mirza}}, \bibinfo {author}
  {\bibfnamefont {C.}~\bibnamefont {Schurgers}}, \bibinfo {author}
  {\bibfnamefont {R.}~\bibnamefont {Kastner}}, \ and\ \bibinfo {author}
  {\bibfnamefont {A.}~\bibnamefont {Boch}},\ }\bibfield  {title} {\enquote
  {\bibinfo {title} {A swarm of autonomous miniature underwater robot drifters
  for exploring submesoscale ocean dynamics},}\ }\href@noop {} {\bibfield
  {journal} {\bibinfo  {journal} {Nature communications}\ }\textbf {\bibinfo
  {volume} {8}},\ \bibinfo {pages} {1--8} (\bibinfo {year} {2017})}\BibitemShut
  {NoStop}%
\bibitem [{\citenamefont {InvenSense}(2013)}]{invensense2013mpu6050}%
  \BibitemOpen
  \bibfield  {author} {\bibinfo {author} {\bibnamefont {InvenSense}},\ }\href
  {https://pdf1.alldatasheet.com/datasheet-pdf/view/1132807/TDK/MPU-6050.html}
  {\emph {\bibinfo {title} {MPU-6000 and MPU-6050 Register Map and Descriptions
  Revision 4.2}}} (\bibinfo {year} {2013}),\ \bibinfo {note} {retrieved from
  InvenSense website}\BibitemShut {NoStop}%
\bibitem [{\citenamefont {Harvey}\ \emph {et~al.}(2019)\citenamefont {Harvey},
  \citenamefont {Baliga}, \citenamefont {Lavoie},\ and\ \citenamefont
  {Altshuler}}]{harvey2019Wing}%
  \BibitemOpen
  \bibfield  {author} {\bibinfo {author} {\bibfnamefont {C.}~\bibnamefont
  {Harvey}}, \bibinfo {author} {\bibfnamefont {V.~B.}\ \bibnamefont {Baliga}},
  \bibinfo {author} {\bibfnamefont {P.}~\bibnamefont {Lavoie}}, \ and\ \bibinfo
  {author} {\bibfnamefont {D.~L.}\ \bibnamefont {Altshuler}},\ }\bibfield
  {title} {\enquote {\bibinfo {title} {Wing morphing allows gulls to modulate
  static pitch stability during gliding},}\ }\href@noop {} {\bibfield
  {journal} {\bibinfo  {journal} {Journal of The Royal Society Interface}\
  }\textbf {\bibinfo {volume} {16}},\ \bibinfo {pages} {20180641} (\bibinfo
  {year} {2019})}\BibitemShut {NoStop}%
\bibitem [{\citenamefont {Candelier}\ \emph {et~al.}(2025)\citenamefont
  {Candelier}, \citenamefont {Gustavsson}, \citenamefont {Sharma},
  \citenamefont {Sundberg}, \citenamefont {Pumir}, \citenamefont {Bagheri},\
  and\ \citenamefont {Mehlig}}]{candelier2025torques}%
  \BibitemOpen
  \bibfield  {author} {\bibinfo {author} {\bibfnamefont {F.}~\bibnamefont
  {Candelier}}, \bibinfo {author} {\bibfnamefont {K.}~\bibnamefont
  {Gustavsson}}, \bibinfo {author} {\bibfnamefont {P.}~\bibnamefont {Sharma}},
  \bibinfo {author} {\bibfnamefont {L.}~\bibnamefont {Sundberg}}, \bibinfo
  {author} {\bibfnamefont {A.}~\bibnamefont {Pumir}}, \bibinfo {author}
  {\bibfnamefont {G.}~\bibnamefont {Bagheri}}, \ and\ \bibinfo {author}
  {\bibfnamefont {B.}~\bibnamefont {Mehlig}},\ }\bibfield  {title} {\enquote
  {\bibinfo {title} {Torques on curved atmospheric fibers},}\ }\href@noop {}
  {\bibfield  {journal} {\bibinfo  {journal} {Phys. Rev. Res.}\ }\textbf
  {\bibinfo {volume} {7}},\ \bibinfo {pages} {013179} (\bibinfo {year}
  {2025})}\BibitemShut {NoStop}%
\bibitem [{\citenamefont {Huseby}\ \emph {et~al.}(2025)\citenamefont {Huseby},
  \citenamefont {Gissinger}, \citenamefont {Candelier}, \citenamefont {Pujara},
  \citenamefont {Verhille}, \citenamefont {Mehlig},\ and\ \citenamefont
  {Voth}}]{huseby2025helical}%
  \BibitemOpen
  \bibfield  {author} {\bibinfo {author} {\bibfnamefont {E.}~\bibnamefont
  {Huseby}}, \bibinfo {author} {\bibfnamefont {J.}~\bibnamefont {Gissinger}},
  \bibinfo {author} {\bibfnamefont {F.}~\bibnamefont {Candelier}}, \bibinfo
  {author} {\bibfnamefont {N.}~\bibnamefont {Pujara}}, \bibinfo {author}
  {\bibfnamefont {G.}~\bibnamefont {Verhille}}, \bibinfo {author}
  {\bibfnamefont {B.}~\bibnamefont {Mehlig}}, \ and\ \bibinfo {author}
  {\bibfnamefont {G.}~\bibnamefont {Voth}},\ }\bibfield  {title} {\enquote
  {\bibinfo {title} {Helical ribbons: Simple chiral sedimentation},}\
  }\href@noop {} {\bibfield  {journal} {\bibinfo  {journal} {Physical Review
  Fluids}\ }\textbf {\bibinfo {volume} {10}},\ \bibinfo {pages} {024101}
  (\bibinfo {year} {2025})}\BibitemShut {NoStop}%
\bibitem [{\citenamefont {Candelier}\ and\ \citenamefont
  {Mehlig}(2016)}]{Candelier2016}%
  \BibitemOpen
  \bibfield  {author} {\bibinfo {author} {\bibfnamefont {F.}~\bibnamefont
  {Candelier}}\ and\ \bibinfo {author} {\bibfnamefont {B.}~\bibnamefont
  {Mehlig}},\ }\bibfield  {title} {\enquote {\bibinfo {title} {Settling of an
  asymmetric dumbbell in a quiescent fluid},}\ }\href@noop {} {\bibfield
  {journal} {\bibinfo  {journal} {Journal of Fluid Mechanics}\ }\textbf
  {\bibinfo {volume} {802}},\ \bibinfo {pages} {174--185} (\bibinfo {year}
  {2016})}\BibitemShut {NoStop}%
\bibitem [{\citenamefont {Roy}\ \emph {et~al.}(2023)\citenamefont {Roy},
  \citenamefont {Kramel}, \citenamefont {Menon}, \citenamefont {Voth},\ and\
  \citenamefont {Koch}}]{roy2023orientation}%
  \BibitemOpen
  \bibfield  {author} {\bibinfo {author} {\bibfnamefont {A.}~\bibnamefont
  {Roy}}, \bibinfo {author} {\bibfnamefont {S.}~\bibnamefont {Kramel}},
  \bibinfo {author} {\bibfnamefont {U.}~\bibnamefont {Menon}}, \bibinfo
  {author} {\bibfnamefont {G.~A.}\ \bibnamefont {Voth}}, \ and\ \bibinfo
  {author} {\bibfnamefont {D.~L.}\ \bibnamefont {Koch}},\ }\bibfield  {title}
  {\enquote {\bibinfo {title} {Orientation of finite {{Reynolds}} number
  anisotropic particles settling in turbulence},}\ }\href@noop {} {\bibfield
  {journal} {\bibinfo  {journal} {Journal of Non-{Newtonian} Fluid Mechanics}\
  }\textbf {\bibinfo {volume} {318}},\ \bibinfo {pages} {105048} (\bibinfo
  {year} {2023})}\BibitemShut {NoStop}%
\bibitem [{\citenamefont {Ravichandran}\ and\ \citenamefont
  {Wettlaufer}(2023)}]{ravichandran2023orientation}%
  \BibitemOpen
  \bibfield  {author} {\bibinfo {author} {\bibfnamefont {S.}~\bibnamefont
  {Ravichandran}}\ and\ \bibinfo {author} {\bibfnamefont {J.~S.}\ \bibnamefont
  {Wettlaufer}},\ }\bibfield  {title} {\enquote {\bibinfo {title} {Orientation
  dynamics of two-dimensional concavo-convex bodies},}\ }\href@noop {}
  {\bibfield  {journal} {\bibinfo  {journal} {Phys. Rev. Fluids}\ }\textbf
  {\bibinfo {volume} {8}},\ \bibinfo {pages} {L062301} (\bibinfo {year}
  {2023})}\BibitemShut {NoStop}%
\bibitem [{\citenamefont {Maches}\ \emph {et~al.}(2024)\citenamefont {Maches},
  \citenamefont {Houssais}, \citenamefont {Sauret},\ and\ \citenamefont
  {Meiburg}}]{maches2024settling}%
  \BibitemOpen
  \bibfield  {author} {\bibinfo {author} {\bibfnamefont {Z.}~\bibnamefont
  {Maches}}, \bibinfo {author} {\bibfnamefont {M.}~\bibnamefont {Houssais}},
  \bibinfo {author} {\bibfnamefont {A.}~\bibnamefont {Sauret}}, \ and\ \bibinfo
  {author} {\bibfnamefont {E.}~\bibnamefont {Meiburg}},\ }\href@noop {}
  {\enquote {\bibinfo {title} {Settling of two rigidly connected spheres},}\ }
  (\bibinfo {year} {2024}),\ \Eprint {http://arxiv.org/abs/2406.10381}
  {arXiv:2406.10381} \BibitemShut {NoStop}%
\end{thebibliography}%
\vfill\eject

\appendix

\setcounter{figure}{0}
\setcounter{table}{0}
\makeatletter
\renewcommand{\thefigure}{A\arabic{figure}}
\renewcommand{\thetable}{A\arabic{table}}
\renewcommand{\theequation}{A\arabic{equation}}

\section{Details in the reinforcement learning}
\label{section:rl}
We implement Double Deep Q-learning~\cite{Hasselt2016} to increase the stability of training. 
In the implementation of deep Q-learning, the feedforward network is constructed with two hidden layers with 16 and 32 neurons, with leaky ReLU functions as activation functions. 
State transitions are stored in a replay memory. A random minibatch of state transitions is sampled for updating the neuron network at a frequency of every 4 state transitions.

During training, the learning rate, $\alpha = \max(0.999^{N_{\rm t}}\alpha_0,\alpha_{\rm min})$, decays from its initial value $\alpha_0$ to the minimal value $\alpha_{\rm min}$, where $N_{\rm t}$ is the number of updates in the deep Q-network $Q(\ve s,\ve a; \ve \theta_t)$, which is a function to evaluate the action $\ve a$ given a state $\ve s$, and $\ve \theta_t$ is the network parameter.
To encourage the exploration in different actions, an $\epsilon$-greedy policy is used during training. The glider chooses random actions with a small probability $\epsilon$, and chooses the action that maximises $Q(\ve s,\ve a; \ve \theta_t)$ otherwise.
The exploration rate, $\epsilon = \max(\epsilon_0 \frac{500-N}{500}, \epsilon_{\rm min})$, decays in every episode from $\epsilon_0$ to the minimal value, $\epsilon_{\rm min}$, where $N$ is the number of episodes. Hyperparameters for training are listed in Table \ref{Table4}.

\begin{figure}[b]
  \includegraphics{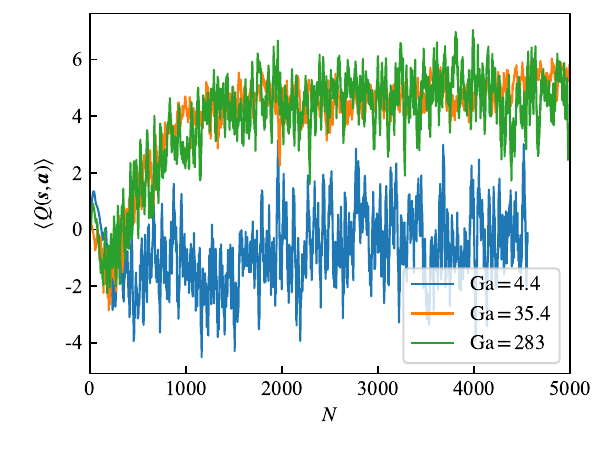} 
  \caption{\label{figS4} Average of value function $\left\langle Q(\ve s, \ve a) \right\rangle$} versus the number of episode $N$. The data is smoothed with a moving average over 20 episodes.
\end{figure}

Figure~\ref{figS4} shows that the training converges at around 4000 episodes.
Here, the training is considered converged when $Q(\ve s, \ve a; \ve \theta_t)$ does not substantially change with $\ve \theta_t$.
This is indicated by that the average of $\left\langle Q(\ve s, \ve a) \right\rangle = N_{\rm t}^{-1} \sum_{i=1}^{N_t} Q(\ve s_i, \ve a_i)$ remains steady as the number of episodes increases.
Upon convergence, the learned strategy is tested without $\epsilon$-greedy exploration, i.e. $\epsilon=0$.

\begin{table}[]
  \setlength{\tabcolsep}{15pt}
  \caption{\label{Table4} Hyperparameters for the deep Q-learning algorithm.}
  \begin{tabular}{lr}
    \hline\hline
    \toprule
    Hyperparameter       & Value \\ \midrule\hline
    Minibatch size       & 256   \\
    Optimizer            & Adam  \\
    Replay memory size   & 8192  \\
    $\gamma$             & 1.0   \\
    $\alpha_0$           & 0.1   \\
    $\alpha_{\rm min}$   & 0.002 \\
    $\epsilon_0$         & 1     \\
    $\epsilon_{\rm min}$ & 0.2   \\
    \bottomrule
    \hline\hline
  \end{tabular}
\end{table}

\setcounter{figure}{0}
\setcounter{table}{0}
\makeatletter
\renewcommand{\thefigure}{B\arabic{figure}}
\renewcommand{\thetable}{B\arabic{table}}
\renewcommand{\theequation}{B\arabic{equation}}

\section{Empirical model for rotation dynamics} \label{sec:empirical_model}

\begin{figure}
  \includegraphics{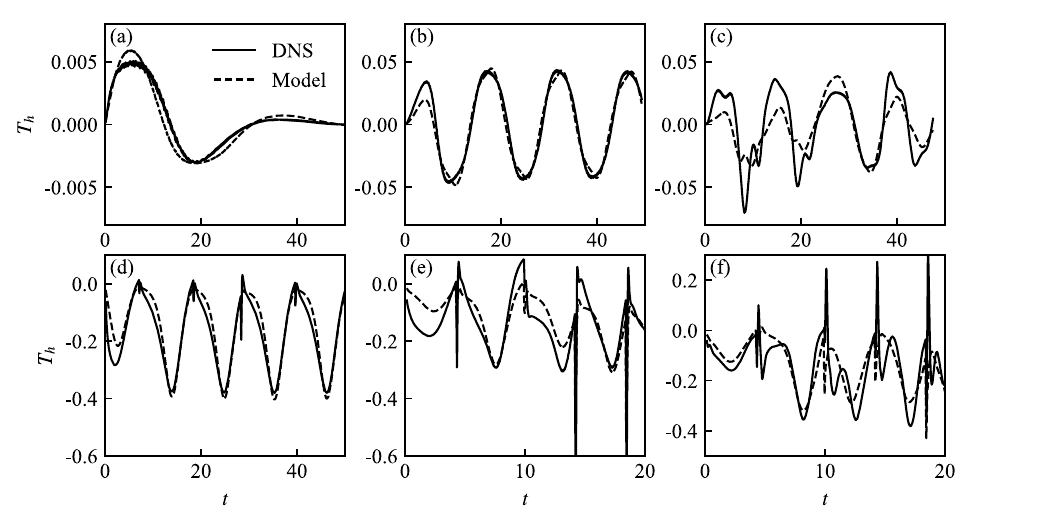} 
  \caption{Comparison of the hydrodynamic torques obtained from DNS and predicted by the empirical model. (a--c) Glider settling with $d=0$ and an initial angle $\theta=\pi/3$. (d--f) Glider following the strategy $d(t)={\rm sgn}[\theta(t)]d_{\rm max}$. (a, d) $\Ga=4.4$; (b, e) $\Ga=35.4$; (c, f) $\Ga=283$.
  \label{fig::fitting}  }
\end{figure}

Figure~\ref{fig::fitting} compares the torques predicted by the model in Eq.~(\ref{eq::torquemodel}) to the torques obtained by our DNS. Fig.~\ref{fig::fitting} (a) to (c) show that the torque on a settling glider with $d=0$ is well predicted by the torque model.
The torque model is less accurate when the glider can change its centre-of-mass according to the heuristic tumbling Eq.~(\ref{eq:tumbling}), as shown in Fig.~\ref{fig::fitting} (d) to (f).
The inaccuracy occurs when an impulsive torque is generated by the movement of the centre-of-mass, as a result of the exchange of angular momentum between fluid and the glider. However, these torque impulses last for a short time, and the torque model predicts the torque well the rest of the time.

\setcounter{figure}{0}
\setcounter{table}{0}
\makeatletter
\renewcommand{\thefigure}{C\arabic{figure}}
\renewcommand{\thetable}{C\arabic{table}}
\renewcommand{\theequation}{C\arabic{equation}}

\section{Importance of signals} \label{section:signals}
\begin{figure}
  \includegraphics{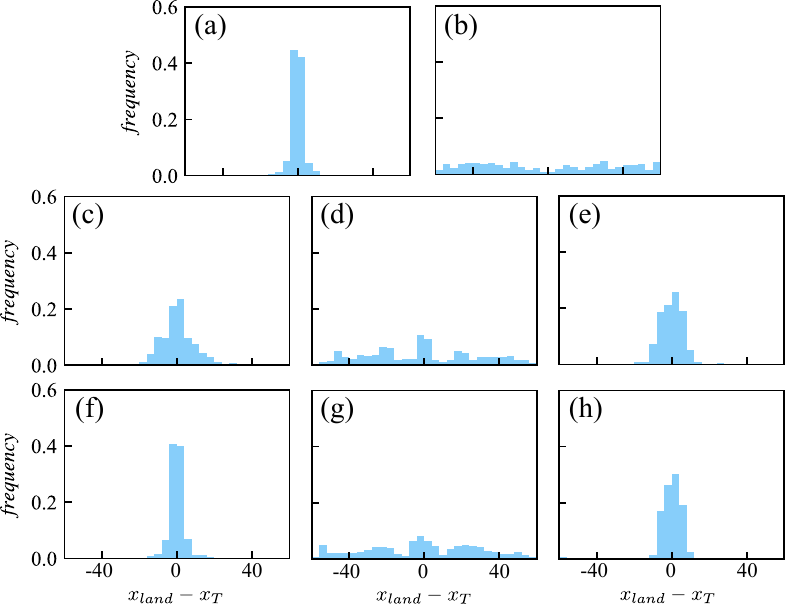} 
  \caption{\label{figS5}  Frequency histograms of errors in landing positions at $\Ga=35.4$. Each panel represents a signal combination. (a) All six signals measured; (b) $x-x_T$ masked; (c) $y-y_T$ masked; (d) $\theta$ masked (e) $v_x$ masked; (f) $v_y$ masked; (g) $\omega$ masked; (h) $v_{x}$ and $v_{y}$ masked.}
\end{figure}

The signals perceived by the glider are $\ve s = \{ x-x_{\rm T}, y-y_{\rm T}, v_x, v_y, \theta, \omega \}$.
To examine their importance, we mask part of them, and run new trainings at $\Ga=35.4$ to see how the performance changes.
For instance, when $x-x_{\rm T}$ is masked, the state consists of the remaining five variables $\ve{s}=\{y-y_{\rm T},v_x,v_y, \theta,\omega\}$.
The performances of strategies with different signals masked are evaluated by the distribution of the errors in landing position, $x_{land}-x_{\rm T}$, as shown in Fig.~\ref{figS5}.
In the reference case, where none of the signal is masked [Fig.~\ref{figS5} (a)], the error is concentrated around zero. However, the error spreads over a wide range when $x-x_{\rm T}$, $\theta $ or $\omega$ is masked, as shown in Fig.~\ref{figS5} (b,d,g), respectively.

\end{document}